%% file: inf1.tex
\documentstyle[12pt]{article}
\input epsf.sty
\newcommand{\be}{\begin{equation}}
\newcommand{\ee}{\end{equation}}
\newcommand{\bea}{\begin{eqnarray}}
\newcommand{\eea}{\end{eqnarray}}

\newcommand{\gm}{\gamma}

\newcommand{\dl}{\delta}

\newcommand{\th}{\theta}

\newcommand{\lm}{\lambda}

\newcommand{\nn}{\nonumber}

\begin{document}
\title{Pre-Inflation
in the Presence of Conformal Coupling}
\author{Apostolos Kuiroukidis\thanks{E-mail:
kouirouki@astro.auth.gr}\\
and\\
Demetrios B. Papadopoulos\\
Department of Physics, \\
\small Section of Astrophysics, Astronomy and Mechanics, \\
\small Aristotle University of Thessaloniki, \\
\small 541 24 Thessaloniki, GREECE\\
}

\maketitle
\begin{abstract}
We consider a massless scalar field, conformally coupled to the Ricci
scalar curvature, in the pre-inflation era of a closed FLRW Universe. The scalar field potential can be of the form of the Coleman-Weinberg one-loop potential, which is flat at the origin and drives the inflationary evolution. For positive values of the conformal parameter $\xi $, less than the critical value $\xi _{c}=(1/6)$, the model admits exact solutions with non-zero minimum scale factor and zero initial Hubble parameter. Thus these solutions can be matched smoothly to the so called Pre-Big-Bang models. At the end of this pre-inflation era one can match inflationary solutions by specifying the form of the potential and the whole solution is of the class $C^{(1)}$.

\end{abstract}

\newpage

\section*{I. Introduction}\
Inflation was historically introduced to solve the problem of monopoles
that could be largely produced in the early Universe. It was subsequently
elaborated into other forms such as the {\it chaotic inflation} scenario
[1-3]. Inflation consists of a short period of accelerated superluminal
expansion of the early Universe, at the end of which the transition to
the standard big bang model should occur [4]. This apparently solves the
problem of the flatness and smoothness of the Universe over such a large
scale of distances [5].

In the scalar field driven inflationary scenario, the matter content
of the Universe has the equation of state of the quantum vacuum,
$P=-\rho $ (with $P, \rho $ the pressure and energy density of matter).
The evolution is regarded as the "rolling" of the value
of the field which is minimally coupled to gravity, in the presence
of the scalar field potential. The form of the potential is motivated
usually
from particle physics arguments. This is effectively a period of
supercooling of the Universe where in fact through a
phase transition the field rolls to its true vacuum state [3,5].

Usually the inflating Universe is taken to have flat spatial sections
\cite{kolb}. However recent WMAP data suggest a Universe with
$\Omega _{tot}=1.02\pm 0.02$ \cite{free}. This implies a closed ($k=+1$)
model. Moreover using Penrose-Hawking-Geroch techniques one can evade
certain theorems (see \cite{tsag} and references therein) and allow for
past-eternally inflating cosmological models that are also singularity
free.

Indeed the problem of the avoidance of the initial singularity and of the exit from inflation are central to all the attempts to construct realistic  inflationary models \cite{tsag}. One of the advantages of the finiteness
of the curvature invariant at the origin and of the zero Hubble parameter,
is that these solutions can be matched smoothly to the so called
Pre-Big-Bang models (see \cite{gasp} and references therein).

In this paper we consider a positively curved FLRW metric, in the presence
of a scalar field. The scalar field is conformally coupled to the spacetime scalar curvature invariant. We are mainly interested in the pre-inflation evolution era. This begins just after the Planck era ($T(GeV)\leq 10^{19}$) and ends at the supposed energy scale that the second order phase transition takes place (e.g., at the GUTs scale $T(GeV)\simeq 2\times 10^{15}$). Then  through the scalar field potential, the field  rolls towards non-zero values, signifying the onset of inflation. In the pre-inflation era we assume that the {\it classical} value of the $\phi$-field remains null, at the origin, ($\phi _{c}=0$) and at the top of the potential profile $V(0)$. The specific form of the potential is not of particular importance here,  since it occurs from various particle physics theories, as long as it is flat enough to drive inflation.

Now we assume that, although the scale factor in this pre-inflation evolution is a classical object, one must consider instead the time evolution of the quantum fluctuations of the $\phi$-field instead. They can be large near the Planck time and as the scale factor increases we have the transition of the Universe towards clearly {\it classical} evolution, so these must be supressed.
This is a reasonable assumption \cite{laz}. The set of coupled equations for the scale factor and the time evolution of the quantum fluctuations, that are suppressed as time evolves, is studied analytically and numerically.
There exist families of solutions where the scale factor has a minimum
non-zero value at the origin and the Hubble parameter is zero. These are particularly interesting because one does not enter in fact the quantum gravity regime at all \cite{tsag}. Moreover they can be matched to the so called Pre-Big-Bang models smoothly \cite{gasp}. Finally at the end of this stage of pre-inflation evolution, the Universe is at an inflationary state (i.e., the second derivative of the scale factor is non-negative
($R^{''}(t)\geq 0$)) and one can match these solutions to classes of inflating solutions, specifying the form of the potential.

\section*{II. Inflation in FLRW Models}\
We consider a Universe described by the metric
\be
ds^{2}=-dt^{2}+R^{2}(t)
\left[\frac{dr^{2}}{1-kr^{2}}+r^{2}(d\th ^{2}+sin^{2}\th d\phi ^{2})\right],
\ee
of the Friedman-Lemaitre-Robertson-Walker (FLRW) class,
where $R(t)$ is the scale factor. We use units where
$\hbar =c=k_{B}=1$ and set $\chi ^{2}:=8\pi G/3$ with \\
$M_{Pl}^{2}=G^{-1}\simeq (1.22\times 10^{19}GeV)^{2}$ [3].\\
Also we assume the presence of a scalar field $\phi $ with energy density and pressure given by
\be
\rho _{\phi }=\frac{\dot{\phi }^{2}}{2}+V(\phi ),\; \; \;
P_{\phi }=\frac{\dot{\phi }^{2}}{2}-V(\phi ),
\ee
where $V(\phi )$ is the potential. In the pre-inflation evolution
the $\phi$-field time evolution is extremely slow. The potential at the origin (i.e. for $\phi =0$) is very flat, as for example in the Coleman-Weinberg
(CW) one-loop, zero temperature case
\be
V(\phi )=V(0)+B\phi ^{4}
\left[ln\left(\frac{\phi ^{2}}{\sigma ^{2}}\right)-\frac{1}{2}\right].
\ee
Here $V(0)=B\sigma ^{4}/2$ with $B\simeq 10^{-3}$ and
$\sigma \simeq 2\times 10^{15}GeV$ the Grand Unification scale.
Also $\chi ^{2}V(0)\simeq (10^{10}GeV)^{2}$ [3].\\
In the pre-inflation era where $10^{15}\leq T(GeV)\leq 10^{19}$,
gravity can be treated clasically, but the other fields must be treated
quantum-mechanically \cite{laz}.
Here we employ the assumption that the $\phi$-field
is a conformally coupled massless scalar field in the action
(see p. 44 of \cite{birr})
\be
S=\int d^{4}x\sqrt{-g}
\left[{\cal R}+6\chi ^{2}\right.
\left(\frac{1}{2}g^{\mu \nu }\phi _{,\mu }\phi _{,\nu }\right.
\left.\left.-\frac{1}{2}\xi {\cal R}(t)\phi ^{2}-V(0)\right)\right].
\ee
It obeys
\be
[\Box +\xi {\cal R}(t)]\phi (x)=0.
\ee
Here ${\cal R}(t)=(6k/R^{2}(t))$ is the Ricci scalar curvature
of the metric in Eq. (1). Usually, in front of the first term in the
action functional, the factor $(1/16\pi G)=(1/6\chi ^{2})$ appears,
which is set to unity by a proper choice of units. However we want the explicit Placnk-scale dependence to be present, so the precise form of the action functional should be the one above. The length scale of our model
is simply $L_{Pl}=t_{Pl}=\chi $. The case $\xi =0$ corresponds to the minimally
coupled scalar field, while $\xi _{c}=(1/6)$ to the conformally
coupled one.\\
The Einstein's equations are exactly given by (see p. 87 of \cite{birr})
\bea
G_{\mu \nu}&+&3\chi ^{2}
\left((1-2\xi )\phi _{;\mu }\phi _{;\nu }+(2\xi-\frac{1}{2})\right.
g_{\mu \nu }g^{\rho \sigma }\phi _{;\rho }\phi _{;\sigma }-\nn \\
&-&2\xi \phi _{;\mu \nu }\phi +2\xi g_{\mu \nu }\phi \Box \phi
\left.-\xi G_{\mu \nu }\phi ^{2}-g_{\mu \nu }V(0)\right)
=3\chi ^{2}T_{\mu \nu }.
\eea
The matter content of our model is given by
$T^{\mu }_{\nu }=diag(\rho ,-P, -P, -P)$ with\\ $P=(\gm -1)\rho $.
The flattness of the potential at the origin effectivelly appears
like a cosmological constant.\\
The equations of motion are (for $k=+1$)
\be
\ddot{\phi }+3H\dot{\phi }+\frac{6\xi }{R^{2}(t)}\phi =0,
\ee
and
\bea
H^{2}+\frac{1}{R^{2}}=\frac{\chi ^{2}}{(1-3\xi \chi ^{2}\phi ^{2})}
\left[\rho +\frac{\dot{\phi }^{2}}{2}+6\xi H\phi \dot{\phi }+V(0)\right],
\\
2\frac{\ddot{R}}{R}+ H^{2}+\frac{1}{R^{2}}=\frac{3\chi
^{2}}{(1-3\xi \chi ^{2}\phi ^{2})} \left[P+(\frac{1}{2}-2\xi
)\frac{\dot{\phi }^{2}}{2}\right. \left.\frac{}{}+4\xi H\phi
\dot{\phi }-V(0)\right].
\eea
Since it is generally accepted that
in this era radiation dominates, we set $\gm =(4/3)$. Eliminating
$\rho $ from Eqs. (8), (9) we obtain
\be
2\frac{\ddot{R}}{R}=
\frac{\chi ^{2}}{(1-3\xi \chi ^{2}\phi ^{2})} [(1-6\xi )\dot{\phi
}^{2}+6\xi H\phi \dot{\phi }-4V(0)].
\ee
We will therefore consider in what follows, Eqs. (7) and (10).
For a recent interesting discussion the reader is referred to
[13].\\
\section*{III. Classes of Solutions from Numerical Integration}
\begin{figure}[h!]
\begin{center}
\input{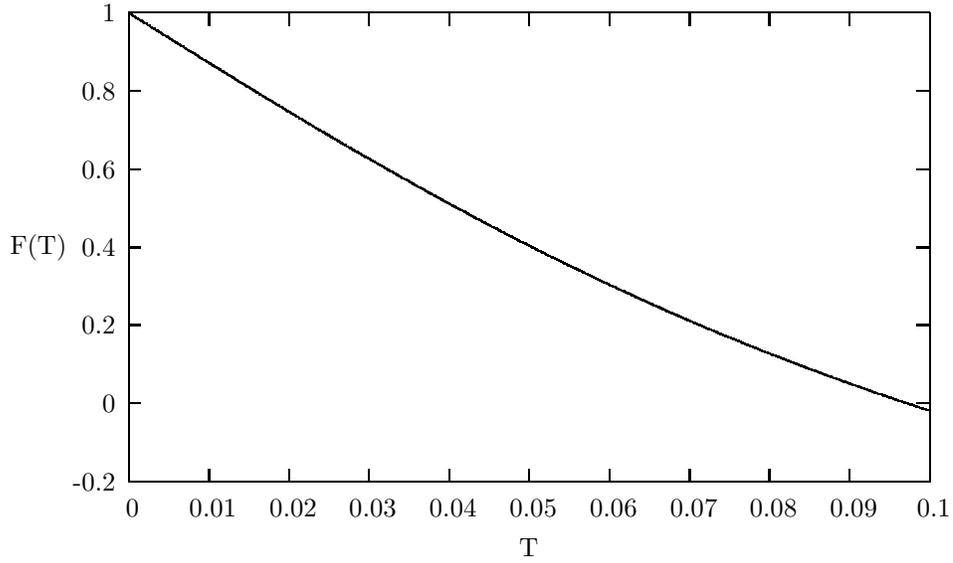}
\end{center}
\caption{The normalized value of the quantum fluctuations, versus the normalized time $T$, of the field
$F(T)$ for $\tilde{\xi }=0.05$. They evolve towards zero value as the
scale factor of the Universe increases.}
\label{ped}
\end{figure}

\begin{figure}[h!]
\begin{center}
\input{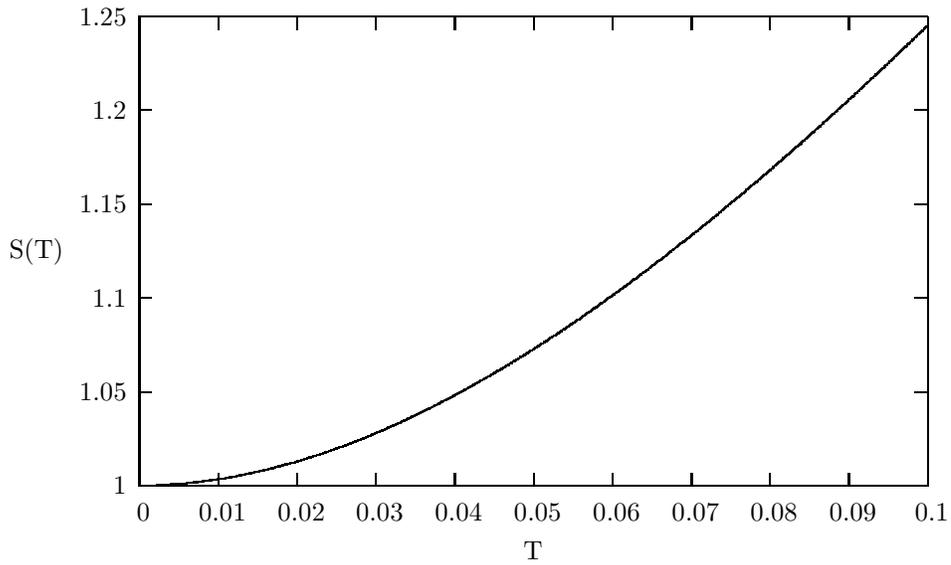}
\end{center}
\caption{The normalized Scale factor $S(T)$, versus the normalized time
$T$, for
$\tilde{\xi }=0.05$. The scale factor assumes a minimum value, with zero initial Hubble parameter.}
\label{scal}
\end{figure}

\
We introduce the parameters $\sigma ,{\cal E}$ through $V(0)=\sigma ^{4}$
and $\chi ^{2}V(0)={\cal E}^{2}$. For example in the case of the
CW potential, $\sigma $ is close to the Grand Unification
scale and ${\cal E}=10^{10}GeV$. We normalize the scale factor
as $(R/R_{min}):=S\geq 1$
because we want a minimum scale factor and also
define $(t/t_{Pl}):=\tilde{t}\geq 1$ because we assume that the evolution takes place past the Planck energy scale. From similar considerations the
combination $F:=\chi \phi $ is dimensionless.
Then Eqs. (7) and (10) give
\bea
& &F^{''}+3\frac{S^{'}}{S}F^{'}+\frac{6\xi }{\lm ^{2}S^{2}}F=0,\\
2\frac{S^{''}}{S}&=&\frac{1}{(1-3\xi F^{2})}\left[(1-6\xi )(F^{'})^{2}
+6\xi \frac{S^{'}}{S}FF^{'}-\chi ^{2}{\cal E}^{2}\right],
\eea
where $\lm :=(R_{min}/\chi )$ and the prime denotes differentiation with respect to the $\tilde{t}$. Also $\dl :=\chi {\cal E}$ is a dimensionless
parameter (for example in the case of the CW potential
\be
\dl :=\chi {\cal E}=\frac{10^{10}GeV}{1.22\times 10^{19}GeV}=
0.82\times 10^{-9}).
\ee
So this parameter, (i.e., $\dl $) which actually stems from the
position of the symmetry breaking energy scale with respect to the Planck
scale, is the first free parameter of our model. It gives the minimum
scale factor in units of the Planck length. The other is of course
$\xi \in [0,\frac{1}{6}]$.\\
Finally rescaling $T:=\dl \tilde{t}$ and $\bar{\lm }:=(\lm \dl )$
we arrive at a completely dimensionless form which is also
esthetically pleasing
\bea
& &F^{''}+3\frac{S^{'}}{S}F^{'}+\frac{6\xi }{\bar{\lm }^{2}S^{2}}F=0,\\
2\frac{S^{''}}{S}&=&\frac{1}{(1-3\xi F^{2})}\left[(1-6\xi )(F^{'})^{2}
+6\xi \frac{S^{'}}{S}FF^{'}-1\right].
\eea
We numerically integrate Eqs. (14) and (15), using as
$\delta \simeq 10^{-9}$. We choose for convenience $\bar{\lm }=1$
which is the third free parameter of our model.
The normalized time begins from $T_{init}=10^{-9}\simeq 0$ i.e.,
just after the Planck time scale, up to $T_{fin}\simeq 0.1$
which is equivalent for the physical time scale $t$, to
the onset of the Grand Unified era. We consider the
time evolution of $F$ as the suppression of the quantum fluctuations
of the $\phi$-field, around its zero classical value ($\phi _{c}=0$),
as time evolves. So the initial value should be
$F_{0}\simeq 1$ and the final $F_{f}\simeq 0$. Then we investigate the
initial value of the field derivative
$F^{'}_{0}=F^{'}_{0}(\xi )$ as a function of
$\xi $ in order that the previous demand is fullfilled. Using double precision numerical integration codes,
we find and plot the Scale factor $S(T)$ also. For $\tilde{\xi }=0.05$ we have the maximum, in absolute value, field derivative $F^{'}_{0}=-12.8$.
The time evolution of the $\phi $-field and the scale
factor for this particular value $\tilde{\xi }=0.05$ are shown in Fig.~\ref{ped} and Fig.~\ref{scal}.
We stress again the fact that we seek solutions with non-zero minimum scale factor and zero initial Hubble parameter. \\
\newpage

\newpage
Of course, every unit of the normalized time $T$ corresponds to
$\dl ^{-1}t_{Pl}\simeq 10^{9}t_{Pl}$ units of the physical time $t$.
We emphasize that this depends on the relative value $V(0)$ with respect to the Planck energy scale.

\begin{figure}[h!]
\begin{center}
\input{ksi}
\end{center}
\caption{The initial time derivative of the field
$F^{'}_{0}=F^{'}_{0}(\xi )$.}
\label{ksou}
\end{figure}
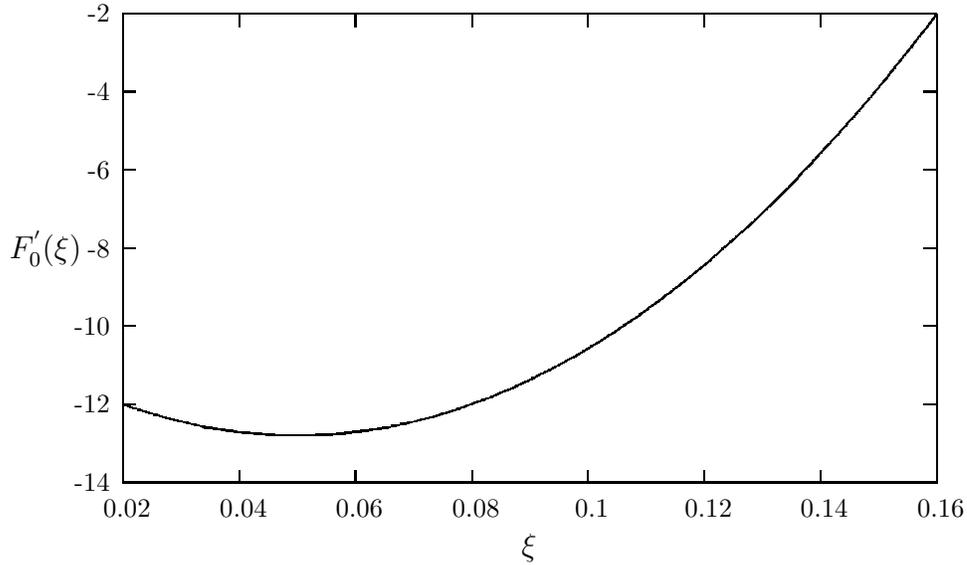

In Fig.~\ref{ksou}. the initial value of the field derivative
$F^{'}_{0}$ versus $\xi $ is plotted for values $0<\xi <(1/6)$.
The numerical integration gives a rather accurate dependence
as $F^{'}_{0}(\xi )=889(\xi -0.05)^{2}-12.8$. \\

\begin{figure}[h!]
\begin{center}
\input{ssi}
\end{center}
\caption{The Dependence $S_{fin}=S_{fin}(\xi )$ of the scale factor
at the end of this period.}
\label{ssou}
\end{figure}
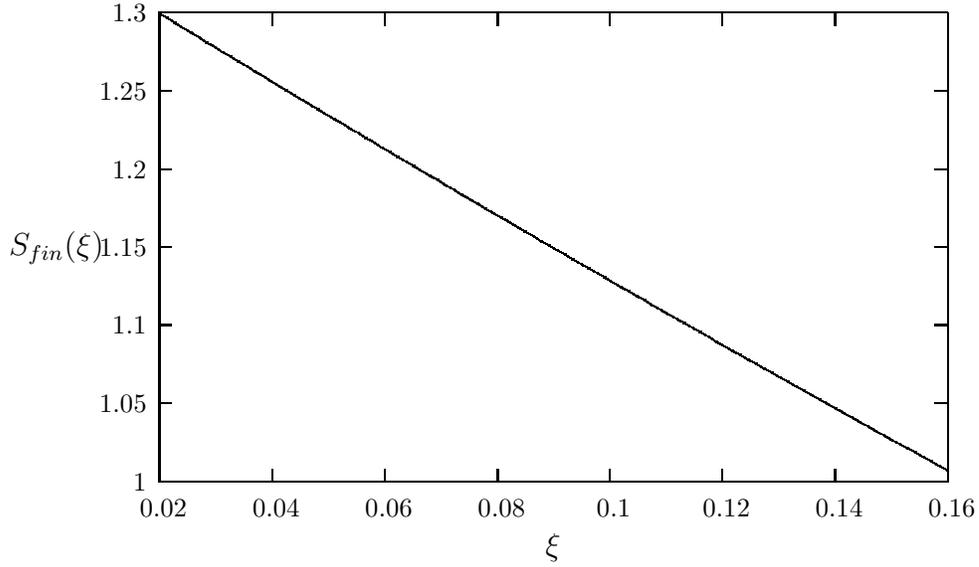

In Fig.~\ref{ssou}. the value of the normalized scale factor
$S_{fin}$ at the time $T\simeq 0.1$,
versus $\xi $ is plotted for values $0<\xi <(1/6)$.
This is effectively the normalized scale factor at the time
the infaltionary era starts. One can therefore match here
inflationary solutions with specific choices of the potential.
The numerical integration gives a rather accurate dependence,
through fitting using Mathematica$^{\textregistered}$, as
$S_{fin}(\xi )=1.344-2.232\xi +0.8\xi ^{2}$. As it is necessary
$S_{fin}>1$.\\

\begin{figure}[h!]
\begin{center}
\input{fder}
\end{center}
\caption{The Dependence $S^{'}_{fin}=S^{'}_{fin}(\xi )$, effectively of the Hubble parameter, at the end of this period.}
\label{fd}
\end{figure}
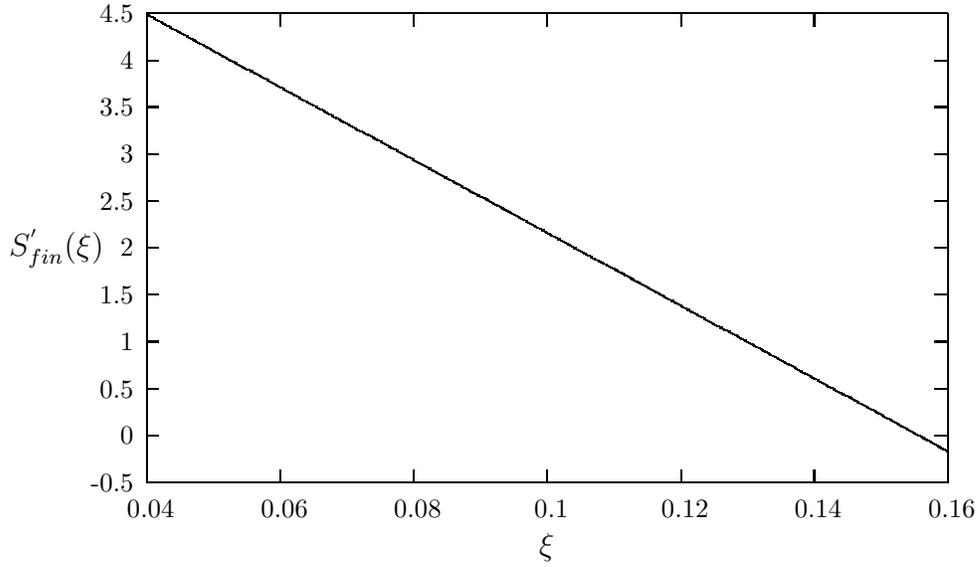

Also the value of the scale factor derivative
$S^{'}_{fin}=S^{'}_{fin}(\xi )$ at the end is
plotted in Fig. ~\ref{fd}. It has the
dependence as $S^{'}_{fin}(\xi )=-38.8\xi +6.043$.
The value of the second derivative $S^{''}_{fin}=S^{''}_{fin}(\xi )$,
as well as of the field time-derivative at the end of this period,
$F^{'}_{fin}=F^{'}_{fin}(\xi )\simeq 0$,
was found to be zero, independent of $\xi $, in the limits of the
numerical accuracy achieved, at the end of the $T$-interval.
So we are exactly at the end of an inflating process and here one
can match inflationary solutions, with a given
value of the Scale factor
and the Hubble parameter, by choosing the appropriate
value of $\xi $.\\
These results imply that the presence of the conformal coupling $\xi $, for a positively curved manifold, act by slowing the inflationary evolution.

\section*{IV. Analytical Considerations}\
The system of equations (14) and (15) has three independent parameters. The  first is $\dl $ which is related to the specific choice of the potential, i.e., the position of $V(0)$ relative to the Planck scale. The second is $\xi $ which depends on the conformal coupling between the scalar curvature and the field. The third is $\bar{\lm }$ which is a free parameter specifying the relation between the minimum scale factor with respect to the Planck length. Once one chooses values for these parameters,
they specify a unique system for Eqs. (14) and (15). Then
there exists a unique initial value for the field  derivative $F^{'}_{0}(\xi )$ that gives the solutions presented. This three-parameter freedom is equivalent to the choice of initial values for the scale factor, the Hubble parameter and the derivative of the scalar field, $\phi ^{'}$ at the end of
this era, $(S_{fin}(\xi ),S^{'}_{fin}(\xi ),F^{'}_{fin}(\xi ))$.\\
The inflationary solution to be matched at this point is
assumed to have $\phi _{0}=0$.

The behaviour of the system of Eqs. (14) and (15) can be approached analytically also.
If we substitute into Eq. (14) the following ansatz (for $\bar{\lm }=1$)
\bea
F(T)&=&1-f_{0}T,\; \; \; f_{0}^{2}=\frac{2}{3},\\
S(T)&=&\sqrt{1+\frac{4\xi }{f_{0}}T-2\xi T^{2}},
\eea
it is satisfied identically. When substituted into Eq. (15) it is an
acceptable solution for small $T$. The parameter $f_{0}$ is $\xi $-invariant but this solution does not satisfy the requirement of zero Hubble parameter for $T=0$. In this solution the scalar field decreases with time $T$, whereas the scale factor {\it increases} in a quasi-linear manner.

We present also a solution of Eqs. (7) and (10) that is not directly relevant to the case discussed here but it is interesting.
We note that an exact solution to Eq. (7) is given by
\bea
R(t)=R_{0}t+R_{min},\; \; \;
\phi (t)=\frac{\phi _{0}t}{(R_{0}t+R_{min})^{3/2}},\; \;
\; R_{0}^{2}=8\xi ,
\eea
which shows that the presence of the conformal coupling
$\xi $ gives the non-zero scale factor evolution $R_{0}$. This is important because it is in general not easy to find simple exact solutions of the Klein-Gordon equation, Eq. (7), for positively curved manifolds. The scalar field increases from zero value as well as the scale factor.\\
When Eq. (18) is substituted into Eq. (10) and one makes the
approximation of small time ($t<<1$), the following relation should
be satisfied
\be
(1-6\xi )\phi _{0}^{2}=4R_{min}^{3}V(0).
\ee
This yields a minimum value $R_{min}$ which is positive for
$\xi \in [0,\frac{1}{6}]$.

\section*{V. Discussion}\
We have considered the pre-inflation era of a positively curved FLRW  Universe
in the presence of the inflaton scalar field $\phi $. This era
begins from the Planck time up to the time where the
symmetry breaking phase transition takes place (e.g., the GUTs scale).
The inflaton field is conformally coupled to the spacetime curvature scalar.

In inflationary theory the quantum fluctuations of scalar fields oscillate until their
wavelength reaches $H^{-1}$. After that they freeze with
amplitude $\dl \phi \sim \frac{H}{2\pi }$ [3,12]. So most of the
wavelengths of the quantum
fluctuations follow the evolution and the only assumption we have made is that their
amplitude is suppressed as time evolves. In Eq. (7), expanding around a
zero classical value for the scalar field, due to the linearity in $\phi $,
one sees that the quantum fluctuations obey the same equation.
We have assumed that they obey also Eq. (10) at least in first
order.\\
As regards the power spectrum of the primordial fluctuations, from Eq. (7)
we see that expanding the scalar field in spherical harmonics, it should be
possible, at least in principle, to detect the possible dependence on the
conformal coupling parameter. This however would require a more detailed
analysis in the framework of a positively curved manifold,
which we defer for a future work.
\footnote{The authors are grateful to the referee for pointing out
this possibility.}

The system has solutions with non-zero {\it minimum scale factor}
and zero initial Hubble parameter.
The quantum fluctuations of the inflaton field are assumed to be
suppressed at the end of this era. This is of considerable importance
because the minimum scale factor is several orders higher than the Planck length scale
(in our case $R_{min}\simeq \dl ^{-1}\chi \simeq 10^{9}L_{Pl}$)
and the Universe does not have to enter the Quantum Gravity era al all [9-11].
Also the finiteness of the
scalar curvature invariant implies that one can continue backwards in time such
solutions, particularly in the context of the so-called Pre-Big-Bang models [9].

As it was described in the previous section a particular inflationary
solution (specifying the form of the scalar potential), with given
initial data for the scale factor $R(t)$, the Hubble parameter $H(t)$,
and the field time derivative $\phi ^{'}(t)$ (because we
assume the initial value of the field to be taken as zero, $\phi (t)=0$)
can be matched to the end of this pre-inflationary evolution. Thus the
whole solution is of the class $C^{(1)}$, i.e., continuously differentiable.

Finally it would be interesting to investigate the role of the possible
conformal coupling between the scalar field and the spacetime curvature
in more general inflationary models, that contain polynomials in the scalar
curvature. Examples of these are the
recently proposed models of [10-11]. Work along this line is in progress.

\end{document}

%% file: ksi.tex
\setlength{\unitlength}{0.240900pt}
\ifx\plotpoint\undefined\newsavebox{\plotpoint}\fi
\sbox{\plotpoint}{\rule[-0.200pt]{0.400pt}{0.400pt}}%
\begin{picture}(1500,900)(0,0)
\font\gnuplot=cmr10 at 10pt
\gnuplot
\sbox{\plotpoint}{\rule[-0.200pt]{0.400pt}{0.400pt}}%
\put(161.0,123.0){\rule[-0.200pt]{4.818pt}{0.400pt}}
\put(141,123){\makebox(0,0)[r]{-14}}
\put(1419.0,123.0){\rule[-0.200pt]{4.818pt}{0.400pt}}
\put(161.0,246.0){\rule[-0.200pt]{4.818pt}{0.400pt}}
\put(141,246){\makebox(0,0)[r]{-12}}
\put(1419.0,246.0){\rule[-0.200pt]{4.818pt}{0.400pt}}
\put(161.0,369.0){\rule[-0.200pt]{4.818pt}{0.400pt}}
\put(141,369){\makebox(0,0)[r]{-10}}
\put(1419.0,369.0){\rule[-0.200pt]{4.818pt}{0.400pt}}
\put(161.0,491.0){\rule[-0.200pt]{4.818pt}{0.400pt}}
\put(141,491){\makebox(0,0)[r]{-8}}
\put(1419.0,491.0){\rule[-0.200pt]{4.818pt}{0.400pt}}
\put(161.0,614.0){\rule[-0.200pt]{4.818pt}{0.400pt}}
\put(141,614){\makebox(0,0)[r]{-6}}
\put(1419.0,614.0){\rule[-0.200pt]{4.818pt}{0.400pt}}
\put(161.0,737.0){\rule[-0.200pt]{4.818pt}{0.400pt}}
\put(141,737){\makebox(0,0)[r]{-4}}
\put(1419.0,737.0){\rule[-0.200pt]{4.818pt}{0.400pt}}
\put(161.0,860.0){\rule[-0.200pt]{4.818pt}{0.400pt}}
\put(141,860){\makebox(0,0)[r]{-2}}
\put(1419.0,860.0){\rule[-0.200pt]{4.818pt}{0.400pt}}
\put(161.0,123.0){\rule[-0.200pt]{0.400pt}{4.818pt}}
\put(161,82){\makebox(0,0){ 0.02}}
\put(161.0,840.0){\rule[-0.200pt]{0.400pt}{4.818pt}}
\put(344.0,123.0){\rule[-0.200pt]{0.400pt}{4.818pt}}
\put(344,82){\makebox(0,0){ 0.04}}
\put(344.0,840.0){\rule[-0.200pt]{0.400pt}{4.818pt}}
\put(526.0,123.0){\rule[-0.200pt]{0.400pt}{4.818pt}}
\put(526,82){\makebox(0,0){ 0.06}}
\put(526.0,840.0){\rule[-0.200pt]{0.400pt}{4.818pt}}
\put(709.0,123.0){\rule[-0.200pt]{0.400pt}{4.818pt}}
\put(709,82){\makebox(0,0){ 0.08}}
\put(709.0,840.0){\rule[-0.200pt]{0.400pt}{4.818pt}}
\put(891.0,123.0){\rule[-0.200pt]{0.400pt}{4.818pt}}
\put(891,82){\makebox(0,0){ 0.1}}
\put(891.0,840.0){\rule[-0.200pt]{0.400pt}{4.818pt}}
\put(1074.0,123.0){\rule[-0.200pt]{0.400pt}{4.818pt}}
\put(1074,82){\makebox(0,0){ 0.12}}
\put(1074.0,840.0){\rule[-0.200pt]{0.400pt}{4.818pt}}
\put(1256.0,123.0){\rule[-0.200pt]{0.400pt}{4.818pt}}
\put(1256,82){\makebox(0,0){ 0.14}}
\put(1256.0,840.0){\rule[-0.200pt]{0.400pt}{4.818pt}}
\put(1439.0,123.0){\rule[-0.200pt]{0.400pt}{4.818pt}}
\put(1439,82){\makebox(0,0){ 0.16}}
\put(1439.0,840.0){\rule[-0.200pt]{0.400pt}{4.818pt}}
\put(161.0,123.0){\rule[-0.200pt]{307.870pt}{0.400pt}}
\put(1439.0,123.0){\rule[-0.200pt]{0.400pt}{177.543pt}}
\put(161.0,860.0){\rule[-0.200pt]{307.870pt}{0.400pt}}
\put(40,491){\makebox(0,0){$F^{'}_{0}(\xi )$}}
\put(800,21){\makebox(0,0){$\xi $}}
\put(161.0,123.0){\rule[-0.200pt]{0.400pt}{177.543pt}}
\put(161,246){\usebox{\plotpoint}}
\multiput(161.00,244.93)(1.378,-0.477){7}{\rule{1.140pt}{0.115pt}}
\multiput(161.00,245.17)(10.634,-5.000){2}{\rule{0.570pt}{0.400pt}}
\multiput(174.00,239.94)(1.797,-0.468){5}{\rule{1.400pt}{0.113pt}}
\multiput(174.00,240.17)(10.094,-4.000){2}{\rule{0.700pt}{0.400pt}}
\multiput(187.00,235.94)(1.797,-0.468){5}{\rule{1.400pt}{0.113pt}}
\multiput(187.00,236.17)(10.094,-4.000){2}{\rule{0.700pt}{0.400pt}}
\multiput(200.00,231.94)(1.797,-0.468){5}{\rule{1.400pt}{0.113pt}}
\multiput(200.00,232.17)(10.094,-4.000){2}{\rule{0.700pt}{0.400pt}}
\multiput(213.00,227.94)(1.797,-0.468){5}{\rule{1.400pt}{0.113pt}}
\multiput(213.00,228.17)(10.094,-4.000){2}{\rule{0.700pt}{0.400pt}}
\multiput(226.00,223.95)(2.472,-0.447){3}{\rule{1.700pt}{0.108pt}}
\multiput(226.00,224.17)(8.472,-3.000){2}{\rule{0.850pt}{0.400pt}}
\multiput(238.00,220.95)(2.695,-0.447){3}{\rule{1.833pt}{0.108pt}}
\multiput(238.00,221.17)(9.195,-3.000){2}{\rule{0.917pt}{0.400pt}}
\multiput(251.00,217.95)(2.695,-0.447){3}{\rule{1.833pt}{0.108pt}}
\multiput(251.00,218.17)(9.195,-3.000){2}{\rule{0.917pt}{0.400pt}}
\multiput(264.00,214.95)(2.695,-0.447){3}{\rule{1.833pt}{0.108pt}}
\multiput(264.00,215.17)(9.195,-3.000){2}{\rule{0.917pt}{0.400pt}}
\multiput(277.00,211.95)(2.695,-0.447){3}{\rule{1.833pt}{0.108pt}}
\multiput(277.00,212.17)(9.195,-3.000){2}{\rule{0.917pt}{0.400pt}}
\put(290,208.17){\rule{2.700pt}{0.400pt}}
\multiput(290.00,209.17)(7.396,-2.000){2}{\rule{1.350pt}{0.400pt}}
\put(303,206.17){\rule{2.700pt}{0.400pt}}
\multiput(303.00,207.17)(7.396,-2.000){2}{\rule{1.350pt}{0.400pt}}
\put(316,204.17){\rule{2.700pt}{0.400pt}}
\multiput(316.00,205.17)(7.396,-2.000){2}{\rule{1.350pt}{0.400pt}}
\put(329,202.17){\rule{2.700pt}{0.400pt}}
\multiput(329.00,203.17)(7.396,-2.000){2}{\rule{1.350pt}{0.400pt}}
\put(342,200.67){\rule{3.132pt}{0.400pt}}
\multiput(342.00,201.17)(6.500,-1.000){2}{\rule{1.566pt}{0.400pt}}
\put(355,199.67){\rule{3.132pt}{0.400pt}}
\multiput(355.00,200.17)(6.500,-1.000){2}{\rule{1.566pt}{0.400pt}}
\put(368,198.67){\rule{2.891pt}{0.400pt}}
\multiput(368.00,199.17)(6.000,-1.000){2}{\rule{1.445pt}{0.400pt}}
\put(380,197.67){\rule{3.132pt}{0.400pt}}
\multiput(380.00,198.17)(6.500,-1.000){2}{\rule{1.566pt}{0.400pt}}
\put(393,196.67){\rule{3.132pt}{0.400pt}}
\multiput(393.00,197.17)(6.500,-1.000){2}{\rule{1.566pt}{0.400pt}}
\put(458,196.67){\rule{3.132pt}{0.400pt}}
\multiput(458.00,196.17)(6.500,1.000){2}{\rule{1.566pt}{0.400pt}}
\put(406.0,197.0){\rule[-0.200pt]{12.527pt}{0.400pt}}
\put(484,197.67){\rule{3.132pt}{0.400pt}}
\multiput(484.00,197.17)(6.500,1.000){2}{\rule{1.566pt}{0.400pt}}
\put(497,198.67){\rule{3.132pt}{0.400pt}}
\multiput(497.00,198.17)(6.500,1.000){2}{\rule{1.566pt}{0.400pt}}
\put(510,200.17){\rule{2.500pt}{0.400pt}}
\multiput(510.00,199.17)(6.811,2.000){2}{\rule{1.250pt}{0.400pt}}
\put(522,201.67){\rule{3.132pt}{0.400pt}}
\multiput(522.00,201.17)(6.500,1.000){2}{\rule{1.566pt}{0.400pt}}
\put(535,203.17){\rule{2.700pt}{0.400pt}}
\multiput(535.00,202.17)(7.396,2.000){2}{\rule{1.350pt}{0.400pt}}
\put(548,205.17){\rule{2.700pt}{0.400pt}}
\multiput(548.00,204.17)(7.396,2.000){2}{\rule{1.350pt}{0.400pt}}
\put(561,207.17){\rule{2.700pt}{0.400pt}}
\multiput(561.00,206.17)(7.396,2.000){2}{\rule{1.350pt}{0.400pt}}
\multiput(574.00,209.61)(2.695,0.447){3}{\rule{1.833pt}{0.108pt}}
\multiput(574.00,208.17)(9.195,3.000){2}{\rule{0.917pt}{0.400pt}}
\multiput(587.00,212.61)(2.695,0.447){3}{\rule{1.833pt}{0.108pt}}
\multiput(587.00,211.17)(9.195,3.000){2}{\rule{0.917pt}{0.400pt}}
\put(600,215.17){\rule{2.700pt}{0.400pt}}
\multiput(600.00,214.17)(7.396,2.000){2}{\rule{1.350pt}{0.400pt}}
\multiput(613.00,217.60)(1.797,0.468){5}{\rule{1.400pt}{0.113pt}}
\multiput(613.00,216.17)(10.094,4.000){2}{\rule{0.700pt}{0.400pt}}
\multiput(626.00,221.61)(2.695,0.447){3}{\rule{1.833pt}{0.108pt}}
\multiput(626.00,220.17)(9.195,3.000){2}{\rule{0.917pt}{0.400pt}}
\multiput(639.00,224.61)(2.695,0.447){3}{\rule{1.833pt}{0.108pt}}
\multiput(639.00,223.17)(9.195,3.000){2}{\rule{0.917pt}{0.400pt}}
\multiput(652.00,227.60)(1.651,0.468){5}{\rule{1.300pt}{0.113pt}}
\multiput(652.00,226.17)(9.302,4.000){2}{\rule{0.650pt}{0.400pt}}
\multiput(664.00,231.60)(1.797,0.468){5}{\rule{1.400pt}{0.113pt}}
\multiput(664.00,230.17)(10.094,4.000){2}{\rule{0.700pt}{0.400pt}}
\multiput(677.00,235.60)(1.797,0.468){5}{\rule{1.400pt}{0.113pt}}
\multiput(677.00,234.17)(10.094,4.000){2}{\rule{0.700pt}{0.400pt}}
\multiput(690.00,239.59)(1.378,0.477){7}{\rule{1.140pt}{0.115pt}}
\multiput(690.00,238.17)(10.634,5.000){2}{\rule{0.570pt}{0.400pt}}
\multiput(703.00,244.59)(1.378,0.477){7}{\rule{1.140pt}{0.115pt}}
\multiput(703.00,243.17)(10.634,5.000){2}{\rule{0.570pt}{0.400pt}}
\multiput(716.00,249.60)(1.797,0.468){5}{\rule{1.400pt}{0.113pt}}
\multiput(716.00,248.17)(10.094,4.000){2}{\rule{0.700pt}{0.400pt}}
\multiput(729.00,253.59)(1.378,0.477){7}{\rule{1.140pt}{0.115pt}}
\multiput(729.00,252.17)(10.634,5.000){2}{\rule{0.570pt}{0.400pt}}
\multiput(742.00,258.59)(1.123,0.482){9}{\rule{0.967pt}{0.116pt}}
\multiput(742.00,257.17)(10.994,6.000){2}{\rule{0.483pt}{0.400pt}}
\multiput(755.00,264.59)(1.378,0.477){7}{\rule{1.140pt}{0.115pt}}
\multiput(755.00,263.17)(10.634,5.000){2}{\rule{0.570pt}{0.400pt}}
\multiput(768.00,269.59)(1.123,0.482){9}{\rule{0.967pt}{0.116pt}}
\multiput(768.00,268.17)(10.994,6.000){2}{\rule{0.483pt}{0.400pt}}
\multiput(781.00,275.59)(1.123,0.482){9}{\rule{0.967pt}{0.116pt}}
\multiput(781.00,274.17)(10.994,6.000){2}{\rule{0.483pt}{0.400pt}}
\multiput(794.00,281.59)(1.033,0.482){9}{\rule{0.900pt}{0.116pt}}
\multiput(794.00,280.17)(10.132,6.000){2}{\rule{0.450pt}{0.400pt}}
\multiput(806.00,287.59)(0.950,0.485){11}{\rule{0.843pt}{0.117pt}}
\multiput(806.00,286.17)(11.251,7.000){2}{\rule{0.421pt}{0.400pt}}
\multiput(819.00,294.59)(1.123,0.482){9}{\rule{0.967pt}{0.116pt}}
\multiput(819.00,293.17)(10.994,6.000){2}{\rule{0.483pt}{0.400pt}}
\multiput(832.00,300.59)(0.950,0.485){11}{\rule{0.843pt}{0.117pt}}
\multiput(832.00,299.17)(11.251,7.000){2}{\rule{0.421pt}{0.400pt}}
\multiput(845.00,307.59)(0.950,0.485){11}{\rule{0.843pt}{0.117pt}}
\multiput(845.00,306.17)(11.251,7.000){2}{\rule{0.421pt}{0.400pt}}
\multiput(858.00,314.59)(0.950,0.485){11}{\rule{0.843pt}{0.117pt}}
\multiput(858.00,313.17)(11.251,7.000){2}{\rule{0.421pt}{0.400pt}}
\multiput(871.00,321.59)(0.824,0.488){13}{\rule{0.750pt}{0.117pt}}
\multiput(871.00,320.17)(11.443,8.000){2}{\rule{0.375pt}{0.400pt}}
\multiput(884.00,329.59)(0.824,0.488){13}{\rule{0.750pt}{0.117pt}}
\multiput(884.00,328.17)(11.443,8.000){2}{\rule{0.375pt}{0.400pt}}
\multiput(897.00,337.59)(0.950,0.485){11}{\rule{0.843pt}{0.117pt}}
\multiput(897.00,336.17)(11.251,7.000){2}{\rule{0.421pt}{0.400pt}}
\multiput(910.00,344.59)(0.728,0.489){15}{\rule{0.678pt}{0.118pt}}
\multiput(910.00,343.17)(11.593,9.000){2}{\rule{0.339pt}{0.400pt}}
\multiput(923.00,353.59)(0.824,0.488){13}{\rule{0.750pt}{0.117pt}}
\multiput(923.00,352.17)(11.443,8.000){2}{\rule{0.375pt}{0.400pt}}
\multiput(936.00,361.59)(0.669,0.489){15}{\rule{0.633pt}{0.118pt}}
\multiput(936.00,360.17)(10.685,9.000){2}{\rule{0.317pt}{0.400pt}}
\multiput(948.00,370.59)(0.824,0.488){13}{\rule{0.750pt}{0.117pt}}
\multiput(948.00,369.17)(11.443,8.000){2}{\rule{0.375pt}{0.400pt}}
\multiput(961.00,378.59)(0.728,0.489){15}{\rule{0.678pt}{0.118pt}}
\multiput(961.00,377.17)(11.593,9.000){2}{\rule{0.339pt}{0.400pt}}
\multiput(974.00,387.58)(0.652,0.491){17}{\rule{0.620pt}{0.118pt}}
\multiput(974.00,386.17)(11.713,10.000){2}{\rule{0.310pt}{0.400pt}}
\multiput(987.00,397.59)(0.728,0.489){15}{\rule{0.678pt}{0.118pt}}
\multiput(987.00,396.17)(11.593,9.000){2}{\rule{0.339pt}{0.400pt}}
\multiput(1000.00,406.58)(0.652,0.491){17}{\rule{0.620pt}{0.118pt}}
\multiput(1000.00,405.17)(11.713,10.000){2}{\rule{0.310pt}{0.400pt}}
\multiput(1013.00,416.58)(0.652,0.491){17}{\rule{0.620pt}{0.118pt}}
\multiput(1013.00,415.17)(11.713,10.000){2}{\rule{0.310pt}{0.400pt}}
\multiput(1026.00,426.58)(0.652,0.491){17}{\rule{0.620pt}{0.118pt}}
\multiput(1026.00,425.17)(11.713,10.000){2}{\rule{0.310pt}{0.400pt}}
\multiput(1039.00,436.58)(0.652,0.491){17}{\rule{0.620pt}{0.118pt}}
\multiput(1039.00,435.17)(11.713,10.000){2}{\rule{0.310pt}{0.400pt}}
\multiput(1052.00,446.58)(0.590,0.492){19}{\rule{0.573pt}{0.118pt}}
\multiput(1052.00,445.17)(11.811,11.000){2}{\rule{0.286pt}{0.400pt}}
\multiput(1065.00,457.58)(0.652,0.491){17}{\rule{0.620pt}{0.118pt}}
\multiput(1065.00,456.17)(11.713,10.000){2}{\rule{0.310pt}{0.400pt}}
\multiput(1078.00,467.58)(0.543,0.492){19}{\rule{0.536pt}{0.118pt}}
\multiput(1078.00,466.17)(10.887,11.000){2}{\rule{0.268pt}{0.400pt}}
\multiput(1090.00,478.58)(0.539,0.492){21}{\rule{0.533pt}{0.119pt}}
\multiput(1090.00,477.17)(11.893,12.000){2}{\rule{0.267pt}{0.400pt}}
\multiput(1103.00,490.58)(0.590,0.492){19}{\rule{0.573pt}{0.118pt}}
\multiput(1103.00,489.17)(11.811,11.000){2}{\rule{0.286pt}{0.400pt}}
\multiput(1116.00,501.58)(0.539,0.492){21}{\rule{0.533pt}{0.119pt}}
\multiput(1116.00,500.17)(11.893,12.000){2}{\rule{0.267pt}{0.400pt}}
\multiput(1129.00,513.58)(0.590,0.492){19}{\rule{0.573pt}{0.118pt}}
\multiput(1129.00,512.17)(11.811,11.000){2}{\rule{0.286pt}{0.400pt}}
\multiput(1142.00,524.58)(0.539,0.492){21}{\rule{0.533pt}{0.119pt}}
\multiput(1142.00,523.17)(11.893,12.000){2}{\rule{0.267pt}{0.400pt}}
\multiput(1155.00,536.58)(0.497,0.493){23}{\rule{0.500pt}{0.119pt}}
\multiput(1155.00,535.17)(11.962,13.000){2}{\rule{0.250pt}{0.400pt}}
\multiput(1168.00,549.58)(0.539,0.492){21}{\rule{0.533pt}{0.119pt}}
\multiput(1168.00,548.17)(11.893,12.000){2}{\rule{0.267pt}{0.400pt}}
\multiput(1181.00,561.58)(0.497,0.493){23}{\rule{0.500pt}{0.119pt}}
\multiput(1181.00,560.17)(11.962,13.000){2}{\rule{0.250pt}{0.400pt}}
\multiput(1194.00,574.58)(0.497,0.493){23}{\rule{0.500pt}{0.119pt}}
\multiput(1194.00,573.17)(11.962,13.000){2}{\rule{0.250pt}{0.400pt}}
\multiput(1207.00,587.58)(0.497,0.493){23}{\rule{0.500pt}{0.119pt}}
\multiput(1207.00,586.17)(11.962,13.000){2}{\rule{0.250pt}{0.400pt}}
\multiput(1220.58,600.00)(0.492,0.582){21}{\rule{0.119pt}{0.567pt}}
\multiput(1219.17,600.00)(12.000,12.824){2}{\rule{0.400pt}{0.283pt}}
\multiput(1232.00,614.58)(0.497,0.493){23}{\rule{0.500pt}{0.119pt}}
\multiput(1232.00,613.17)(11.962,13.000){2}{\rule{0.250pt}{0.400pt}}
\multiput(1245.58,627.00)(0.493,0.536){23}{\rule{0.119pt}{0.531pt}}
\multiput(1244.17,627.00)(13.000,12.898){2}{\rule{0.400pt}{0.265pt}}
\multiput(1258.58,641.00)(0.493,0.536){23}{\rule{0.119pt}{0.531pt}}
\multiput(1257.17,641.00)(13.000,12.898){2}{\rule{0.400pt}{0.265pt}}
\multiput(1271.58,655.00)(0.493,0.536){23}{\rule{0.119pt}{0.531pt}}
\multiput(1270.17,655.00)(13.000,12.898){2}{\rule{0.400pt}{0.265pt}}
\multiput(1284.58,669.00)(0.493,0.576){23}{\rule{0.119pt}{0.562pt}}
\multiput(1283.17,669.00)(13.000,13.834){2}{\rule{0.400pt}{0.281pt}}
\multiput(1297.58,684.00)(0.493,0.536){23}{\rule{0.119pt}{0.531pt}}
\multiput(1296.17,684.00)(13.000,12.898){2}{\rule{0.400pt}{0.265pt}}
\multiput(1310.58,698.00)(0.493,0.576){23}{\rule{0.119pt}{0.562pt}}
\multiput(1309.17,698.00)(13.000,13.834){2}{\rule{0.400pt}{0.281pt}}
\multiput(1323.58,713.00)(0.493,0.576){23}{\rule{0.119pt}{0.562pt}}
\multiput(1322.17,713.00)(13.000,13.834){2}{\rule{0.400pt}{0.281pt}}
\multiput(1336.58,728.00)(0.493,0.616){23}{\rule{0.119pt}{0.592pt}}
\multiput(1335.17,728.00)(13.000,14.771){2}{\rule{0.400pt}{0.296pt}}
\multiput(1349.58,744.00)(0.493,0.576){23}{\rule{0.119pt}{0.562pt}}
\multiput(1348.17,744.00)(13.000,13.834){2}{\rule{0.400pt}{0.281pt}}
\multiput(1362.58,759.00)(0.492,0.669){21}{\rule{0.119pt}{0.633pt}}
\multiput(1361.17,759.00)(12.000,14.685){2}{\rule{0.400pt}{0.317pt}}
\multiput(1374.58,775.00)(0.493,0.616){23}{\rule{0.119pt}{0.592pt}}
\multiput(1373.17,775.00)(13.000,14.771){2}{\rule{0.400pt}{0.296pt}}
\multiput(1387.58,791.00)(0.493,0.616){23}{\rule{0.119pt}{0.592pt}}
\multiput(1386.17,791.00)(13.000,14.771){2}{\rule{0.400pt}{0.296pt}}
\multiput(1400.58,807.00)(0.493,0.655){23}{\rule{0.119pt}{0.623pt}}
\multiput(1399.17,807.00)(13.000,15.707){2}{\rule{0.400pt}{0.312pt}}
\multiput(1413.58,824.00)(0.493,0.616){23}{\rule{0.119pt}{0.592pt}}
\multiput(1412.17,824.00)(13.000,14.771){2}{\rule{0.400pt}{0.296pt}}
\multiput(1426.58,840.00)(0.493,0.655){23}{\rule{0.119pt}{0.623pt}}
\multiput(1425.17,840.00)(13.000,15.707){2}{\rule{0.400pt}{0.312pt}}
\put(471.0,198.0){\rule[-0.200pt]{3.132pt}{0.400pt}}
\end{picture}

%% file: ssi.tex
\setlength{\unitlength}{0.240900pt}
\ifx\plotpoint\undefined\newsavebox{\plotpoint}\fi
\begin{picture}(1500,900)(0,0)
\font\gnuplot=cmr10 at 10pt
\gnuplot
\sbox{\plotpoint}{\rule[-0.200pt]{0.400pt}{0.400pt}}%
\put(201.0,123.0){\rule[-0.200pt]{4.818pt}{0.400pt}}
\put(181,123){\makebox(0,0)[r]{ 1}}
\put(1419.0,123.0){\rule[-0.200pt]{4.818pt}{0.400pt}}
\put(201.0,246.0){\rule[-0.200pt]{4.818pt}{0.400pt}}
\put(181,246){\makebox(0,0)[r]{ 1.05}}
\put(1419.0,246.0){\rule[-0.200pt]{4.818pt}{0.400pt}}
\put(201.0,369.0){\rule[-0.200pt]{4.818pt}{0.400pt}}
\put(181,369){\makebox(0,0)[r]{ 1.1}}
\put(1419.0,369.0){\rule[-0.200pt]{4.818pt}{0.400pt}}
\put(201.0,492.0){\rule[-0.200pt]{4.818pt}{0.400pt}}
\put(181,492){\makebox(0,0)[r]{ 1.15}}
\put(1419.0,492.0){\rule[-0.200pt]{4.818pt}{0.400pt}}
\put(201.0,614.0){\rule[-0.200pt]{4.818pt}{0.400pt}}
\put(181,614){\makebox(0,0)[r]{ 1.2}}
\put(1419.0,614.0){\rule[-0.200pt]{4.818pt}{0.400pt}}
\put(201.0,737.0){\rule[-0.200pt]{4.818pt}{0.400pt}}
\put(181,737){\makebox(0,0)[r]{ 1.25}}
\put(1419.0,737.0){\rule[-0.200pt]{4.818pt}{0.400pt}}
\put(201.0,860.0){\rule[-0.200pt]{4.818pt}{0.400pt}}
\put(181,860){\makebox(0,0)[r]{ 1.3}}
\put(1419.0,860.0){\rule[-0.200pt]{4.818pt}{0.400pt}}
\put(201.0,123.0){\rule[-0.200pt]{0.400pt}{4.818pt}}
\put(201,82){\makebox(0,0){ 0.02}}
\put(201.0,840.0){\rule[-0.200pt]{0.400pt}{4.818pt}}
\put(378.0,123.0){\rule[-0.200pt]{0.400pt}{4.818pt}}
\put(378,82){\makebox(0,0){ 0.04}}
\put(378.0,840.0){\rule[-0.200pt]{0.400pt}{4.818pt}}
\put(555.0,123.0){\rule[-0.200pt]{0.400pt}{4.818pt}}
\put(555,82){\makebox(0,0){ 0.06}}
\put(555.0,840.0){\rule[-0.200pt]{0.400pt}{4.818pt}}
\put(732.0,123.0){\rule[-0.200pt]{0.400pt}{4.818pt}}
\put(732,82){\makebox(0,0){ 0.08}}
\put(732.0,840.0){\rule[-0.200pt]{0.400pt}{4.818pt}}
\put(908.0,123.0){\rule[-0.200pt]{0.400pt}{4.818pt}}
\put(908,82){\makebox(0,0){ 0.1}}
\put(908.0,840.0){\rule[-0.200pt]{0.400pt}{4.818pt}}
\put(1085.0,123.0){\rule[-0.200pt]{0.400pt}{4.818pt}}
\put(1085,82){\makebox(0,0){ 0.12}}
\put(1085.0,840.0){\rule[-0.200pt]{0.400pt}{4.818pt}}
\put(1262.0,123.0){\rule[-0.200pt]{0.400pt}{4.818pt}}
\put(1262,82){\makebox(0,0){ 0.14}}
\put(1262.0,840.0){\rule[-0.200pt]{0.400pt}{4.818pt}}
\put(1439.0,123.0){\rule[-0.200pt]{0.400pt}{4.818pt}}
\put(1439,82){\makebox(0,0){ 0.16}}
\put(1439.0,840.0){\rule[-0.200pt]{0.400pt}{4.818pt}}
\put(201.0,123.0){\rule[-0.200pt]{298.234pt}{0.400pt}}
\put(1439.0,123.0){\rule[-0.200pt]{0.400pt}{177.543pt}}
\put(201.0,860.0){\rule[-0.200pt]{298.234pt}{0.400pt}}
\put(40,491){\makebox(0,0){$S_{fin}(\xi )$}}
\put(820,21){\makebox(0,0){$\xi $}}
\put(201.0,123.0){\rule[-0.200pt]{0.400pt}{177.543pt}}
\put(201,859){\usebox{\plotpoint}}
\multiput(201.00,857.93)(0.950,-0.485){11}{\rule{0.843pt}{0.117pt}}
\multiput(201.00,858.17)(11.251,-7.000){2}{\rule{0.421pt}{0.400pt}}
\multiput(214.00,850.93)(0.758,-0.488){13}{\rule{0.700pt}{0.117pt}}
\multiput(214.00,851.17)(10.547,-8.000){2}{\rule{0.350pt}{0.400pt}}
\multiput(226.00,842.93)(0.824,-0.488){13}{\rule{0.750pt}{0.117pt}}
\multiput(226.00,843.17)(11.443,-8.000){2}{\rule{0.375pt}{0.400pt}}
\multiput(239.00,834.93)(0.874,-0.485){11}{\rule{0.786pt}{0.117pt}}
\multiput(239.00,835.17)(10.369,-7.000){2}{\rule{0.393pt}{0.400pt}}
\multiput(251.00,827.93)(0.824,-0.488){13}{\rule{0.750pt}{0.117pt}}
\multiput(251.00,828.17)(11.443,-8.000){2}{\rule{0.375pt}{0.400pt}}
\multiput(264.00,819.93)(0.758,-0.488){13}{\rule{0.700pt}{0.117pt}}
\multiput(264.00,820.17)(10.547,-8.000){2}{\rule{0.350pt}{0.400pt}}
\multiput(276.00,811.93)(0.950,-0.485){11}{\rule{0.843pt}{0.117pt}}
\multiput(276.00,812.17)(11.251,-7.000){2}{\rule{0.421pt}{0.400pt}}
\multiput(289.00,804.93)(0.758,-0.488){13}{\rule{0.700pt}{0.117pt}}
\multiput(289.00,805.17)(10.547,-8.000){2}{\rule{0.350pt}{0.400pt}}
\multiput(301.00,796.93)(0.950,-0.485){11}{\rule{0.843pt}{0.117pt}}
\multiput(301.00,797.17)(11.251,-7.000){2}{\rule{0.421pt}{0.400pt}}
\multiput(314.00,789.93)(0.758,-0.488){13}{\rule{0.700pt}{0.117pt}}
\multiput(314.00,790.17)(10.547,-8.000){2}{\rule{0.350pt}{0.400pt}}
\multiput(326.00,781.93)(0.950,-0.485){11}{\rule{0.843pt}{0.117pt}}
\multiput(326.00,782.17)(11.251,-7.000){2}{\rule{0.421pt}{0.400pt}}
\multiput(339.00,774.93)(0.758,-0.488){13}{\rule{0.700pt}{0.117pt}}
\multiput(339.00,775.17)(10.547,-8.000){2}{\rule{0.350pt}{0.400pt}}
\multiput(351.00,766.93)(0.950,-0.485){11}{\rule{0.843pt}{0.117pt}}
\multiput(351.00,767.17)(11.251,-7.000){2}{\rule{0.421pt}{0.400pt}}
\multiput(364.00,759.93)(0.758,-0.488){13}{\rule{0.700pt}{0.117pt}}
\multiput(364.00,760.17)(10.547,-8.000){2}{\rule{0.350pt}{0.400pt}}
\multiput(376.00,751.93)(0.824,-0.488){13}{\rule{0.750pt}{0.117pt}}
\multiput(376.00,752.17)(11.443,-8.000){2}{\rule{0.375pt}{0.400pt}}
\multiput(389.00,743.93)(0.874,-0.485){11}{\rule{0.786pt}{0.117pt}}
\multiput(389.00,744.17)(10.369,-7.000){2}{\rule{0.393pt}{0.400pt}}
\multiput(401.00,736.93)(0.824,-0.488){13}{\rule{0.750pt}{0.117pt}}
\multiput(401.00,737.17)(11.443,-8.000){2}{\rule{0.375pt}{0.400pt}}
\multiput(414.00,728.93)(0.874,-0.485){11}{\rule{0.786pt}{0.117pt}}
\multiput(414.00,729.17)(10.369,-7.000){2}{\rule{0.393pt}{0.400pt}}
\multiput(426.00,721.93)(0.824,-0.488){13}{\rule{0.750pt}{0.117pt}}
\multiput(426.00,722.17)(11.443,-8.000){2}{\rule{0.375pt}{0.400pt}}
\multiput(439.00,713.93)(0.874,-0.485){11}{\rule{0.786pt}{0.117pt}}
\multiput(439.00,714.17)(10.369,-7.000){2}{\rule{0.393pt}{0.400pt}}
\multiput(451.00,706.93)(0.824,-0.488){13}{\rule{0.750pt}{0.117pt}}
\multiput(451.00,707.17)(11.443,-8.000){2}{\rule{0.375pt}{0.400pt}}
\multiput(464.00,698.93)(0.874,-0.485){11}{\rule{0.786pt}{0.117pt}}
\multiput(464.00,699.17)(10.369,-7.000){2}{\rule{0.393pt}{0.400pt}}
\multiput(476.00,691.93)(0.950,-0.485){11}{\rule{0.843pt}{0.117pt}}
\multiput(476.00,692.17)(11.251,-7.000){2}{\rule{0.421pt}{0.400pt}}
\multiput(489.00,684.93)(0.758,-0.488){13}{\rule{0.700pt}{0.117pt}}
\multiput(489.00,685.17)(10.547,-8.000){2}{\rule{0.350pt}{0.400pt}}
\multiput(501.00,676.93)(0.950,-0.485){11}{\rule{0.843pt}{0.117pt}}
\multiput(501.00,677.17)(11.251,-7.000){2}{\rule{0.421pt}{0.400pt}}
\multiput(514.00,669.93)(0.758,-0.488){13}{\rule{0.700pt}{0.117pt}}
\multiput(514.00,670.17)(10.547,-8.000){2}{\rule{0.350pt}{0.400pt}}
\multiput(526.00,661.93)(0.950,-0.485){11}{\rule{0.843pt}{0.117pt}}
\multiput(526.00,662.17)(11.251,-7.000){2}{\rule{0.421pt}{0.400pt}}
\multiput(539.00,654.93)(0.758,-0.488){13}{\rule{0.700pt}{0.117pt}}
\multiput(539.00,655.17)(10.547,-8.000){2}{\rule{0.350pt}{0.400pt}}
\multiput(551.00,646.93)(0.950,-0.485){11}{\rule{0.843pt}{0.117pt}}
\multiput(551.00,647.17)(11.251,-7.000){2}{\rule{0.421pt}{0.400pt}}
\multiput(564.00,639.93)(0.758,-0.488){13}{\rule{0.700pt}{0.117pt}}
\multiput(564.00,640.17)(10.547,-8.000){2}{\rule{0.350pt}{0.400pt}}
\multiput(576.00,631.93)(0.950,-0.485){11}{\rule{0.843pt}{0.117pt}}
\multiput(576.00,632.17)(11.251,-7.000){2}{\rule{0.421pt}{0.400pt}}
\multiput(589.00,624.93)(0.874,-0.485){11}{\rule{0.786pt}{0.117pt}}
\multiput(589.00,625.17)(10.369,-7.000){2}{\rule{0.393pt}{0.400pt}}
\multiput(601.00,617.93)(0.824,-0.488){13}{\rule{0.750pt}{0.117pt}}
\multiput(601.00,618.17)(11.443,-8.000){2}{\rule{0.375pt}{0.400pt}}
\multiput(614.00,609.93)(0.874,-0.485){11}{\rule{0.786pt}{0.117pt}}
\multiput(614.00,610.17)(10.369,-7.000){2}{\rule{0.393pt}{0.400pt}}
\multiput(626.00,602.93)(0.950,-0.485){11}{\rule{0.843pt}{0.117pt}}
\multiput(626.00,603.17)(11.251,-7.000){2}{\rule{0.421pt}{0.400pt}}
\multiput(639.00,595.93)(0.758,-0.488){13}{\rule{0.700pt}{0.117pt}}
\multiput(639.00,596.17)(10.547,-8.000){2}{\rule{0.350pt}{0.400pt}}
\multiput(651.00,587.93)(0.950,-0.485){11}{\rule{0.843pt}{0.117pt}}
\multiput(651.00,588.17)(11.251,-7.000){2}{\rule{0.421pt}{0.400pt}}
\multiput(664.00,580.93)(0.758,-0.488){13}{\rule{0.700pt}{0.117pt}}
\multiput(664.00,581.17)(10.547,-8.000){2}{\rule{0.350pt}{0.400pt}}
\multiput(676.00,572.93)(0.950,-0.485){11}{\rule{0.843pt}{0.117pt}}
\multiput(676.00,573.17)(11.251,-7.000){2}{\rule{0.421pt}{0.400pt}}
\multiput(689.00,565.93)(0.874,-0.485){11}{\rule{0.786pt}{0.117pt}}
\multiput(689.00,566.17)(10.369,-7.000){2}{\rule{0.393pt}{0.400pt}}
\multiput(701.00,558.93)(0.824,-0.488){13}{\rule{0.750pt}{0.117pt}}
\multiput(701.00,559.17)(11.443,-8.000){2}{\rule{0.375pt}{0.400pt}}
\multiput(714.00,550.93)(0.874,-0.485){11}{\rule{0.786pt}{0.117pt}}
\multiput(714.00,551.17)(10.369,-7.000){2}{\rule{0.393pt}{0.400pt}}
\multiput(726.00,543.93)(0.950,-0.485){11}{\rule{0.843pt}{0.117pt}}
\multiput(726.00,544.17)(11.251,-7.000){2}{\rule{0.421pt}{0.400pt}}
\multiput(739.00,536.93)(0.874,-0.485){11}{\rule{0.786pt}{0.117pt}}
\multiput(739.00,537.17)(10.369,-7.000){2}{\rule{0.393pt}{0.400pt}}
\multiput(751.00,529.93)(0.824,-0.488){13}{\rule{0.750pt}{0.117pt}}
\multiput(751.00,530.17)(11.443,-8.000){2}{\rule{0.375pt}{0.400pt}}
\multiput(764.00,521.93)(0.874,-0.485){11}{\rule{0.786pt}{0.117pt}}
\multiput(764.00,522.17)(10.369,-7.000){2}{\rule{0.393pt}{0.400pt}}
\multiput(776.00,514.93)(0.950,-0.485){11}{\rule{0.843pt}{0.117pt}}
\multiput(776.00,515.17)(11.251,-7.000){2}{\rule{0.421pt}{0.400pt}}
\multiput(789.00,507.93)(0.758,-0.488){13}{\rule{0.700pt}{0.117pt}}
\multiput(789.00,508.17)(10.547,-8.000){2}{\rule{0.350pt}{0.400pt}}
\multiput(801.00,499.93)(0.950,-0.485){11}{\rule{0.843pt}{0.117pt}}
\multiput(801.00,500.17)(11.251,-7.000){2}{\rule{0.421pt}{0.400pt}}
\multiput(814.00,492.93)(0.874,-0.485){11}{\rule{0.786pt}{0.117pt}}
\multiput(814.00,493.17)(10.369,-7.000){2}{\rule{0.393pt}{0.400pt}}
\multiput(826.00,485.93)(0.950,-0.485){11}{\rule{0.843pt}{0.117pt}}
\multiput(826.00,486.17)(11.251,-7.000){2}{\rule{0.421pt}{0.400pt}}
\multiput(839.00,478.93)(0.758,-0.488){13}{\rule{0.700pt}{0.117pt}}
\multiput(839.00,479.17)(10.547,-8.000){2}{\rule{0.350pt}{0.400pt}}
\multiput(851.00,470.93)(0.950,-0.485){11}{\rule{0.843pt}{0.117pt}}
\multiput(851.00,471.17)(11.251,-7.000){2}{\rule{0.421pt}{0.400pt}}
\multiput(864.00,463.93)(0.874,-0.485){11}{\rule{0.786pt}{0.117pt}}
\multiput(864.00,464.17)(10.369,-7.000){2}{\rule{0.393pt}{0.400pt}}
\multiput(876.00,456.93)(0.950,-0.485){11}{\rule{0.843pt}{0.117pt}}
\multiput(876.00,457.17)(11.251,-7.000){2}{\rule{0.421pt}{0.400pt}}
\multiput(889.00,449.93)(0.874,-0.485){11}{\rule{0.786pt}{0.117pt}}
\multiput(889.00,450.17)(10.369,-7.000){2}{\rule{0.393pt}{0.400pt}}
\multiput(901.00,442.93)(0.824,-0.488){13}{\rule{0.750pt}{0.117pt}}
\multiput(901.00,443.17)(11.443,-8.000){2}{\rule{0.375pt}{0.400pt}}
\multiput(914.00,434.93)(0.874,-0.485){11}{\rule{0.786pt}{0.117pt}}
\multiput(914.00,435.17)(10.369,-7.000){2}{\rule{0.393pt}{0.400pt}}
\multiput(926.00,427.93)(0.950,-0.485){11}{\rule{0.843pt}{0.117pt}}
\multiput(926.00,428.17)(11.251,-7.000){2}{\rule{0.421pt}{0.400pt}}
\multiput(939.00,420.93)(0.874,-0.485){11}{\rule{0.786pt}{0.117pt}}
\multiput(939.00,421.17)(10.369,-7.000){2}{\rule{0.393pt}{0.400pt}}
\multiput(951.00,413.93)(0.950,-0.485){11}{\rule{0.843pt}{0.117pt}}
\multiput(951.00,414.17)(11.251,-7.000){2}{\rule{0.421pt}{0.400pt}}
\multiput(964.00,406.93)(0.758,-0.488){13}{\rule{0.700pt}{0.117pt}}
\multiput(964.00,407.17)(10.547,-8.000){2}{\rule{0.350pt}{0.400pt}}
\multiput(976.00,398.93)(0.950,-0.485){11}{\rule{0.843pt}{0.117pt}}
\multiput(976.00,399.17)(11.251,-7.000){2}{\rule{0.421pt}{0.400pt}}
\multiput(989.00,391.93)(0.874,-0.485){11}{\rule{0.786pt}{0.117pt}}
\multiput(989.00,392.17)(10.369,-7.000){2}{\rule{0.393pt}{0.400pt}}
\multiput(1001.00,384.93)(0.950,-0.485){11}{\rule{0.843pt}{0.117pt}}
\multiput(1001.00,385.17)(11.251,-7.000){2}{\rule{0.421pt}{0.400pt}}
\multiput(1014.00,377.93)(0.874,-0.485){11}{\rule{0.786pt}{0.117pt}}
\multiput(1014.00,378.17)(10.369,-7.000){2}{\rule{0.393pt}{0.400pt}}
\multiput(1026.00,370.93)(0.950,-0.485){11}{\rule{0.843pt}{0.117pt}}
\multiput(1026.00,371.17)(11.251,-7.000){2}{\rule{0.421pt}{0.400pt}}
\multiput(1039.00,363.93)(0.874,-0.485){11}{\rule{0.786pt}{0.117pt}}
\multiput(1039.00,364.17)(10.369,-7.000){2}{\rule{0.393pt}{0.400pt}}
\multiput(1051.00,356.93)(0.950,-0.485){11}{\rule{0.843pt}{0.117pt}}
\multiput(1051.00,357.17)(11.251,-7.000){2}{\rule{0.421pt}{0.400pt}}
\multiput(1064.00,349.93)(0.758,-0.488){13}{\rule{0.700pt}{0.117pt}}
\multiput(1064.00,350.17)(10.547,-8.000){2}{\rule{0.350pt}{0.400pt}}
\multiput(1076.00,341.93)(0.950,-0.485){11}{\rule{0.843pt}{0.117pt}}
\multiput(1076.00,342.17)(11.251,-7.000){2}{\rule{0.421pt}{0.400pt}}
\multiput(1089.00,334.93)(0.874,-0.485){11}{\rule{0.786pt}{0.117pt}}
\multiput(1089.00,335.17)(10.369,-7.000){2}{\rule{0.393pt}{0.400pt}}
\multiput(1101.00,327.93)(0.950,-0.485){11}{\rule{0.843pt}{0.117pt}}
\multiput(1101.00,328.17)(11.251,-7.000){2}{\rule{0.421pt}{0.400pt}}
\multiput(1114.00,320.93)(0.874,-0.485){11}{\rule{0.786pt}{0.117pt}}
\multiput(1114.00,321.17)(10.369,-7.000){2}{\rule{0.393pt}{0.400pt}}
\multiput(1126.00,313.93)(0.950,-0.485){11}{\rule{0.843pt}{0.117pt}}
\multiput(1126.00,314.17)(11.251,-7.000){2}{\rule{0.421pt}{0.400pt}}
\multiput(1139.00,306.93)(0.874,-0.485){11}{\rule{0.786pt}{0.117pt}}
\multiput(1139.00,307.17)(10.369,-7.000){2}{\rule{0.393pt}{0.400pt}}
\multiput(1151.00,299.93)(0.950,-0.485){11}{\rule{0.843pt}{0.117pt}}
\multiput(1151.00,300.17)(11.251,-7.000){2}{\rule{0.421pt}{0.400pt}}
\multiput(1164.00,292.93)(0.874,-0.485){11}{\rule{0.786pt}{0.117pt}}
\multiput(1164.00,293.17)(10.369,-7.000){2}{\rule{0.393pt}{0.400pt}}
\multiput(1176.00,285.93)(0.950,-0.485){11}{\rule{0.843pt}{0.117pt}}
\multiput(1176.00,286.17)(11.251,-7.000){2}{\rule{0.421pt}{0.400pt}}
\multiput(1189.00,278.93)(0.874,-0.485){11}{\rule{0.786pt}{0.117pt}}
\multiput(1189.00,279.17)(10.369,-7.000){2}{\rule{0.393pt}{0.400pt}}
\multiput(1201.00,271.93)(0.950,-0.485){11}{\rule{0.843pt}{0.117pt}}
\multiput(1201.00,272.17)(11.251,-7.000){2}{\rule{0.421pt}{0.400pt}}
\multiput(1214.00,264.93)(0.874,-0.485){11}{\rule{0.786pt}{0.117pt}}
\multiput(1214.00,265.17)(10.369,-7.000){2}{\rule{0.393pt}{0.400pt}}
\multiput(1226.00,257.93)(0.950,-0.485){11}{\rule{0.843pt}{0.117pt}}
\multiput(1226.00,258.17)(11.251,-7.000){2}{\rule{0.421pt}{0.400pt}}
\multiput(1239.00,250.93)(0.874,-0.485){11}{\rule{0.786pt}{0.117pt}}
\multiput(1239.00,251.17)(10.369,-7.000){2}{\rule{0.393pt}{0.400pt}}
\multiput(1251.00,243.93)(0.950,-0.485){11}{\rule{0.843pt}{0.117pt}}
\multiput(1251.00,244.17)(11.251,-7.000){2}{\rule{0.421pt}{0.400pt}}
\multiput(1264.00,236.93)(0.874,-0.485){11}{\rule{0.786pt}{0.117pt}}
\multiput(1264.00,237.17)(10.369,-7.000){2}{\rule{0.393pt}{0.400pt}}
\multiput(1276.00,229.93)(0.950,-0.485){11}{\rule{0.843pt}{0.117pt}}
\multiput(1276.00,230.17)(11.251,-7.000){2}{\rule{0.421pt}{0.400pt}}
\multiput(1289.00,222.93)(0.874,-0.485){11}{\rule{0.786pt}{0.117pt}}
\multiput(1289.00,223.17)(10.369,-7.000){2}{\rule{0.393pt}{0.400pt}}
\multiput(1301.00,215.93)(0.950,-0.485){11}{\rule{0.843pt}{0.117pt}}
\multiput(1301.00,216.17)(11.251,-7.000){2}{\rule{0.421pt}{0.400pt}}
\multiput(1314.00,208.93)(0.874,-0.485){11}{\rule{0.786pt}{0.117pt}}
\multiput(1314.00,209.17)(10.369,-7.000){2}{\rule{0.393pt}{0.400pt}}
\multiput(1326.00,201.93)(0.950,-0.485){11}{\rule{0.843pt}{0.117pt}}
\multiput(1326.00,202.17)(11.251,-7.000){2}{\rule{0.421pt}{0.400pt}}
\multiput(1339.00,194.93)(0.874,-0.485){11}{\rule{0.786pt}{0.117pt}}
\multiput(1339.00,195.17)(10.369,-7.000){2}{\rule{0.393pt}{0.400pt}}
\multiput(1351.00,187.93)(0.950,-0.485){11}{\rule{0.843pt}{0.117pt}}
\multiput(1351.00,188.17)(11.251,-7.000){2}{\rule{0.421pt}{0.400pt}}
\multiput(1364.00,180.93)(1.033,-0.482){9}{\rule{0.900pt}{0.116pt}}
\multiput(1364.00,181.17)(10.132,-6.000){2}{\rule{0.450pt}{0.400pt}}
\multiput(1376.00,174.93)(0.950,-0.485){11}{\rule{0.843pt}{0.117pt}}
\multiput(1376.00,175.17)(11.251,-7.000){2}{\rule{0.421pt}{0.400pt}}
\multiput(1389.00,167.93)(0.874,-0.485){11}{\rule{0.786pt}{0.117pt}}
\multiput(1389.00,168.17)(10.369,-7.000){2}{\rule{0.393pt}{0.400pt}}
\multiput(1401.00,160.93)(0.950,-0.485){11}{\rule{0.843pt}{0.117pt}}
\multiput(1401.00,161.17)(11.251,-7.000){2}{\rule{0.421pt}{0.400pt}}
\multiput(1414.00,153.93)(0.874,-0.485){11}{\rule{0.786pt}{0.117pt}}
\multiput(1414.00,154.17)(10.369,-7.000){2}{\rule{0.393pt}{0.400pt}}
\multiput(1426.00,146.93)(0.950,-0.485){11}{\rule{0.843pt}{0.117pt}}
\multiput(1426.00,147.17)(11.251,-7.000){2}{\rule{0.421pt}{0.400pt}}
\end{picture}

%% file: fder.tex
\setlength{\unitlength}{0.240900pt}
\ifx\plotpoint\undefined\newsavebox{\plotpoint}\fi
\sbox{\plotpoint}{\rule[-0.200pt]{0.400pt}{0.400pt}}%
\begin{picture}(1500,900)(0,0)
\font\gnuplot=cmr10 at 10pt
\gnuplot
\sbox{\plotpoint}{\rule[-0.200pt]{0.400pt}{0.400pt}}%
\put(181.0,123.0){\rule[-0.200pt]{4.818pt}{0.400pt}}
\put(161,123){\makebox(0,0)[r]{-0.5}}
\put(1419.0,123.0){\rule[-0.200pt]{4.818pt}{0.400pt}}
\put(181.0,197.0){\rule[-0.200pt]{4.818pt}{0.400pt}}
\put(161,197){\makebox(0,0)[r]{ 0}}
\put(1419.0,197.0){\rule[-0.200pt]{4.818pt}{0.400pt}}
\put(181.0,270.0){\rule[-0.200pt]{4.818pt}{0.400pt}}
\put(161,270){\makebox(0,0)[r]{ 0.5}}
\put(1419.0,270.0){\rule[-0.200pt]{4.818pt}{0.400pt}}
\put(181.0,344.0){\rule[-0.200pt]{4.818pt}{0.400pt}}
\put(161,344){\makebox(0,0)[r]{ 1}}
\put(1419.0,344.0){\rule[-0.200pt]{4.818pt}{0.400pt}}
\put(181.0,418.0){\rule[-0.200pt]{4.818pt}{0.400pt}}
\put(161,418){\makebox(0,0)[r]{ 1.5}}
\put(1419.0,418.0){\rule[-0.200pt]{4.818pt}{0.400pt}}
\put(181.0,492.0){\rule[-0.200pt]{4.818pt}{0.400pt}}
\put(161,492){\makebox(0,0)[r]{ 2}}
\put(1419.0,492.0){\rule[-0.200pt]{4.818pt}{0.400pt}}
\put(181.0,565.0){\rule[-0.200pt]{4.818pt}{0.400pt}}
\put(161,565){\makebox(0,0)[r]{ 2.5}}
\put(1419.0,565.0){\rule[-0.200pt]{4.818pt}{0.400pt}}
\put(181.0,639.0){\rule[-0.200pt]{4.818pt}{0.400pt}}
\put(161,639){\makebox(0,0)[r]{ 3}}
\put(1419.0,639.0){\rule[-0.200pt]{4.818pt}{0.400pt}}
\put(181.0,713.0){\rule[-0.200pt]{4.818pt}{0.400pt}}
\put(161,713){\makebox(0,0)[r]{ 3.5}}
\put(1419.0,713.0){\rule[-0.200pt]{4.818pt}{0.400pt}}
\put(181.0,786.0){\rule[-0.200pt]{4.818pt}{0.400pt}}
\put(161,786){\makebox(0,0)[r]{ 4}}
\put(1419.0,786.0){\rule[-0.200pt]{4.818pt}{0.400pt}}
\put(181.0,860.0){\rule[-0.200pt]{4.818pt}{0.400pt}}
\put(161,860){\makebox(0,0)[r]{ 4.5}}
\put(1419.0,860.0){\rule[-0.200pt]{4.818pt}{0.400pt}}
\put(181.0,123.0){\rule[-0.200pt]{0.400pt}{4.818pt}}
\put(181,82){\makebox(0,0){ 0.04}}
\put(181.0,840.0){\rule[-0.200pt]{0.400pt}{4.818pt}}
\put(391.0,123.0){\rule[-0.200pt]{0.400pt}{4.818pt}}
\put(391,82){\makebox(0,0){ 0.06}}
\put(391.0,840.0){\rule[-0.200pt]{0.400pt}{4.818pt}}
\put(600.0,123.0){\rule[-0.200pt]{0.400pt}{4.818pt}}
\put(600,82){\makebox(0,0){ 0.08}}
\put(600.0,840.0){\rule[-0.200pt]{0.400pt}{4.818pt}}
\put(810.0,123.0){\rule[-0.200pt]{0.400pt}{4.818pt}}
\put(810,82){\makebox(0,0){ 0.1}}
\put(810.0,840.0){\rule[-0.200pt]{0.400pt}{4.818pt}}
\put(1020.0,123.0){\rule[-0.200pt]{0.400pt}{4.818pt}}
\put(1020,82){\makebox(0,0){ 0.12}}
\put(1020.0,840.0){\rule[-0.200pt]{0.400pt}{4.818pt}}
\put(1229.0,123.0){\rule[-0.200pt]{0.400pt}{4.818pt}}
\put(1229,82){\makebox(0,0){ 0.14}}
\put(1229.0,840.0){\rule[-0.200pt]{0.400pt}{4.818pt}}
\put(1439.0,123.0){\rule[-0.200pt]{0.400pt}{4.818pt}}
\put(1439,82){\makebox(0,0){ 0.16}}
\put(1439.0,840.0){\rule[-0.200pt]{0.400pt}{4.818pt}}
\put(181.0,123.0){\rule[-0.200pt]{303.052pt}{0.400pt}}
\put(1439.0,123.0){\rule[-0.200pt]{0.400pt}{177.543pt}}
\put(181.0,860.0){\rule[-0.200pt]{303.052pt}{0.400pt}}
\put(40,491){\makebox(0,0){$S^{'}_{fin}(\xi )$}}
\put(810,21){\makebox(0,0){$\xi $}}
\put(181.0,123.0){\rule[-0.200pt]{0.400pt}{177.543pt}}
\put(181,859){\usebox{\plotpoint}}
\multiput(181.00,857.93)(0.950,-0.485){11}{\rule{0.843pt}{0.117pt}}
\multiput(181.00,858.17)(11.251,-7.000){2}{\rule{0.421pt}{0.400pt}}
\multiput(194.00,850.93)(0.874,-0.485){11}{\rule{0.786pt}{0.117pt}}
\multiput(194.00,851.17)(10.369,-7.000){2}{\rule{0.393pt}{0.400pt}}
\multiput(206.00,843.93)(0.950,-0.485){11}{\rule{0.843pt}{0.117pt}}
\multiput(206.00,844.17)(11.251,-7.000){2}{\rule{0.421pt}{0.400pt}}
\multiput(219.00,836.93)(0.950,-0.485){11}{\rule{0.843pt}{0.117pt}}
\multiput(219.00,837.17)(11.251,-7.000){2}{\rule{0.421pt}{0.400pt}}
\multiput(232.00,829.93)(0.950,-0.485){11}{\rule{0.843pt}{0.117pt}}
\multiput(232.00,830.17)(11.251,-7.000){2}{\rule{0.421pt}{0.400pt}}
\multiput(245.00,822.93)(0.874,-0.485){11}{\rule{0.786pt}{0.117pt}}
\multiput(245.00,823.17)(10.369,-7.000){2}{\rule{0.393pt}{0.400pt}}
\multiput(257.00,815.93)(0.950,-0.485){11}{\rule{0.843pt}{0.117pt}}
\multiput(257.00,816.17)(11.251,-7.000){2}{\rule{0.421pt}{0.400pt}}
\multiput(270.00,808.93)(0.950,-0.485){11}{\rule{0.843pt}{0.117pt}}
\multiput(270.00,809.17)(11.251,-7.000){2}{\rule{0.421pt}{0.400pt}}
\multiput(283.00,801.93)(0.874,-0.485){11}{\rule{0.786pt}{0.117pt}}
\multiput(283.00,802.17)(10.369,-7.000){2}{\rule{0.393pt}{0.400pt}}
\multiput(295.00,794.93)(0.950,-0.485){11}{\rule{0.843pt}{0.117pt}}
\multiput(295.00,795.17)(11.251,-7.000){2}{\rule{0.421pt}{0.400pt}}
\multiput(308.00,787.93)(0.950,-0.485){11}{\rule{0.843pt}{0.117pt}}
\multiput(308.00,788.17)(11.251,-7.000){2}{\rule{0.421pt}{0.400pt}}
\multiput(321.00,780.93)(0.874,-0.485){11}{\rule{0.786pt}{0.117pt}}
\multiput(321.00,781.17)(10.369,-7.000){2}{\rule{0.393pt}{0.400pt}}
\multiput(333.00,773.93)(1.123,-0.482){9}{\rule{0.967pt}{0.116pt}}
\multiput(333.00,774.17)(10.994,-6.000){2}{\rule{0.483pt}{0.400pt}}
\multiput(346.00,767.93)(0.950,-0.485){11}{\rule{0.843pt}{0.117pt}}
\multiput(346.00,768.17)(11.251,-7.000){2}{\rule{0.421pt}{0.400pt}}
\multiput(359.00,760.93)(0.950,-0.485){11}{\rule{0.843pt}{0.117pt}}
\multiput(359.00,761.17)(11.251,-7.000){2}{\rule{0.421pt}{0.400pt}}
\multiput(372.00,753.93)(0.874,-0.485){11}{\rule{0.786pt}{0.117pt}}
\multiput(372.00,754.17)(10.369,-7.000){2}{\rule{0.393pt}{0.400pt}}
\multiput(384.00,746.93)(0.950,-0.485){11}{\rule{0.843pt}{0.117pt}}
\multiput(384.00,747.17)(11.251,-7.000){2}{\rule{0.421pt}{0.400pt}}
\multiput(397.00,739.93)(0.950,-0.485){11}{\rule{0.843pt}{0.117pt}}
\multiput(397.00,740.17)(11.251,-7.000){2}{\rule{0.421pt}{0.400pt}}
\multiput(410.00,732.93)(0.874,-0.485){11}{\rule{0.786pt}{0.117pt}}
\multiput(410.00,733.17)(10.369,-7.000){2}{\rule{0.393pt}{0.400pt}}
\multiput(422.00,725.93)(0.950,-0.485){11}{\rule{0.843pt}{0.117pt}}
\multiput(422.00,726.17)(11.251,-7.000){2}{\rule{0.421pt}{0.400pt}}
\multiput(435.00,718.93)(0.950,-0.485){11}{\rule{0.843pt}{0.117pt}}
\multiput(435.00,719.17)(11.251,-7.000){2}{\rule{0.421pt}{0.400pt}}
\multiput(448.00,711.93)(0.950,-0.485){11}{\rule{0.843pt}{0.117pt}}
\multiput(448.00,712.17)(11.251,-7.000){2}{\rule{0.421pt}{0.400pt}}
\multiput(461.00,704.93)(0.874,-0.485){11}{\rule{0.786pt}{0.117pt}}
\multiput(461.00,705.17)(10.369,-7.000){2}{\rule{0.393pt}{0.400pt}}
\multiput(473.00,697.93)(0.950,-0.485){11}{\rule{0.843pt}{0.117pt}}
\multiput(473.00,698.17)(11.251,-7.000){2}{\rule{0.421pt}{0.400pt}}
\multiput(486.00,690.93)(0.950,-0.485){11}{\rule{0.843pt}{0.117pt}}
\multiput(486.00,691.17)(11.251,-7.000){2}{\rule{0.421pt}{0.400pt}}
\multiput(499.00,683.93)(0.874,-0.485){11}{\rule{0.786pt}{0.117pt}}
\multiput(499.00,684.17)(10.369,-7.000){2}{\rule{0.393pt}{0.400pt}}
\multiput(511.00,676.93)(1.123,-0.482){9}{\rule{0.967pt}{0.116pt}}
\multiput(511.00,677.17)(10.994,-6.000){2}{\rule{0.483pt}{0.400pt}}
\multiput(524.00,670.93)(0.950,-0.485){11}{\rule{0.843pt}{0.117pt}}
\multiput(524.00,671.17)(11.251,-7.000){2}{\rule{0.421pt}{0.400pt}}
\multiput(537.00,663.93)(0.950,-0.485){11}{\rule{0.843pt}{0.117pt}}
\multiput(537.00,664.17)(11.251,-7.000){2}{\rule{0.421pt}{0.400pt}}
\multiput(550.00,656.93)(0.874,-0.485){11}{\rule{0.786pt}{0.117pt}}
\multiput(550.00,657.17)(10.369,-7.000){2}{\rule{0.393pt}{0.400pt}}
\multiput(562.00,649.93)(0.950,-0.485){11}{\rule{0.843pt}{0.117pt}}
\multiput(562.00,650.17)(11.251,-7.000){2}{\rule{0.421pt}{0.400pt}}
\multiput(575.00,642.93)(0.950,-0.485){11}{\rule{0.843pt}{0.117pt}}
\multiput(575.00,643.17)(11.251,-7.000){2}{\rule{0.421pt}{0.400pt}}
\multiput(588.00,635.93)(0.874,-0.485){11}{\rule{0.786pt}{0.117pt}}
\multiput(588.00,636.17)(10.369,-7.000){2}{\rule{0.393pt}{0.400pt}}
\multiput(600.00,628.93)(0.950,-0.485){11}{\rule{0.843pt}{0.117pt}}
\multiput(600.00,629.17)(11.251,-7.000){2}{\rule{0.421pt}{0.400pt}}
\multiput(613.00,621.93)(0.950,-0.485){11}{\rule{0.843pt}{0.117pt}}
\multiput(613.00,622.17)(11.251,-7.000){2}{\rule{0.421pt}{0.400pt}}
\multiput(626.00,614.93)(0.874,-0.485){11}{\rule{0.786pt}{0.117pt}}
\multiput(626.00,615.17)(10.369,-7.000){2}{\rule{0.393pt}{0.400pt}}
\multiput(638.00,607.93)(0.950,-0.485){11}{\rule{0.843pt}{0.117pt}}
\multiput(638.00,608.17)(11.251,-7.000){2}{\rule{0.421pt}{0.400pt}}
\multiput(651.00,600.93)(0.950,-0.485){11}{\rule{0.843pt}{0.117pt}}
\multiput(651.00,601.17)(11.251,-7.000){2}{\rule{0.421pt}{0.400pt}}
\multiput(664.00,593.93)(0.950,-0.485){11}{\rule{0.843pt}{0.117pt}}
\multiput(664.00,594.17)(11.251,-7.000){2}{\rule{0.421pt}{0.400pt}}
\multiput(677.00,586.93)(0.874,-0.485){11}{\rule{0.786pt}{0.117pt}}
\multiput(677.00,587.17)(10.369,-7.000){2}{\rule{0.393pt}{0.400pt}}
\multiput(689.00,579.93)(0.950,-0.485){11}{\rule{0.843pt}{0.117pt}}
\multiput(689.00,580.17)(11.251,-7.000){2}{\rule{0.421pt}{0.400pt}}
\multiput(702.00,572.93)(1.123,-0.482){9}{\rule{0.967pt}{0.116pt}}
\multiput(702.00,573.17)(10.994,-6.000){2}{\rule{0.483pt}{0.400pt}}
\multiput(715.00,566.93)(0.874,-0.485){11}{\rule{0.786pt}{0.117pt}}
\multiput(715.00,567.17)(10.369,-7.000){2}{\rule{0.393pt}{0.400pt}}
\multiput(727.00,559.93)(0.950,-0.485){11}{\rule{0.843pt}{0.117pt}}
\multiput(727.00,560.17)(11.251,-7.000){2}{\rule{0.421pt}{0.400pt}}
\multiput(740.00,552.93)(0.950,-0.485){11}{\rule{0.843pt}{0.117pt}}
\multiput(740.00,553.17)(11.251,-7.000){2}{\rule{0.421pt}{0.400pt}}
\multiput(753.00,545.93)(0.950,-0.485){11}{\rule{0.843pt}{0.117pt}}
\multiput(753.00,546.17)(11.251,-7.000){2}{\rule{0.421pt}{0.400pt}}
\multiput(766.00,538.93)(0.874,-0.485){11}{\rule{0.786pt}{0.117pt}}
\multiput(766.00,539.17)(10.369,-7.000){2}{\rule{0.393pt}{0.400pt}}
\multiput(778.00,531.93)(0.950,-0.485){11}{\rule{0.843pt}{0.117pt}}
\multiput(778.00,532.17)(11.251,-7.000){2}{\rule{0.421pt}{0.400pt}}
\multiput(791.00,524.93)(0.950,-0.485){11}{\rule{0.843pt}{0.117pt}}
\multiput(791.00,525.17)(11.251,-7.000){2}{\rule{0.421pt}{0.400pt}}
\multiput(804.00,517.93)(0.874,-0.485){11}{\rule{0.786pt}{0.117pt}}
\multiput(804.00,518.17)(10.369,-7.000){2}{\rule{0.393pt}{0.400pt}}
\multiput(816.00,510.93)(0.950,-0.485){11}{\rule{0.843pt}{0.117pt}}
\multiput(816.00,511.17)(11.251,-7.000){2}{\rule{0.421pt}{0.400pt}}
\multiput(829.00,503.93)(0.950,-0.485){11}{\rule{0.843pt}{0.117pt}}
\multiput(829.00,504.17)(11.251,-7.000){2}{\rule{0.421pt}{0.400pt}}
\multiput(842.00,496.93)(0.874,-0.485){11}{\rule{0.786pt}{0.117pt}}
\multiput(842.00,497.17)(10.369,-7.000){2}{\rule{0.393pt}{0.400pt}}
\multiput(854.00,489.93)(0.950,-0.485){11}{\rule{0.843pt}{0.117pt}}
\multiput(854.00,490.17)(11.251,-7.000){2}{\rule{0.421pt}{0.400pt}}
\multiput(867.00,482.93)(0.950,-0.485){11}{\rule{0.843pt}{0.117pt}}
\multiput(867.00,483.17)(11.251,-7.000){2}{\rule{0.421pt}{0.400pt}}
\multiput(880.00,475.93)(0.950,-0.485){11}{\rule{0.843pt}{0.117pt}}
\multiput(880.00,476.17)(11.251,-7.000){2}{\rule{0.421pt}{0.400pt}}
\multiput(893.00,468.93)(1.033,-0.482){9}{\rule{0.900pt}{0.116pt}}
\multiput(893.00,469.17)(10.132,-6.000){2}{\rule{0.450pt}{0.400pt}}
\multiput(905.00,462.93)(0.950,-0.485){11}{\rule{0.843pt}{0.117pt}}
\multiput(905.00,463.17)(11.251,-7.000){2}{\rule{0.421pt}{0.400pt}}
\multiput(918.00,455.93)(0.950,-0.485){11}{\rule{0.843pt}{0.117pt}}
\multiput(918.00,456.17)(11.251,-7.000){2}{\rule{0.421pt}{0.400pt}}
\multiput(931.00,448.93)(0.874,-0.485){11}{\rule{0.786pt}{0.117pt}}
\multiput(931.00,449.17)(10.369,-7.000){2}{\rule{0.393pt}{0.400pt}}
\multiput(943.00,441.93)(0.950,-0.485){11}{\rule{0.843pt}{0.117pt}}
\multiput(943.00,442.17)(11.251,-7.000){2}{\rule{0.421pt}{0.400pt}}
\multiput(956.00,434.93)(0.950,-0.485){11}{\rule{0.843pt}{0.117pt}}
\multiput(956.00,435.17)(11.251,-7.000){2}{\rule{0.421pt}{0.400pt}}
\multiput(969.00,427.93)(0.950,-0.485){11}{\rule{0.843pt}{0.117pt}}
\multiput(969.00,428.17)(11.251,-7.000){2}{\rule{0.421pt}{0.400pt}}
\multiput(982.00,420.93)(0.874,-0.485){11}{\rule{0.786pt}{0.117pt}}
\multiput(982.00,421.17)(10.369,-7.000){2}{\rule{0.393pt}{0.400pt}}
\multiput(994.00,413.93)(0.950,-0.485){11}{\rule{0.843pt}{0.117pt}}
\multiput(994.00,414.17)(11.251,-7.000){2}{\rule{0.421pt}{0.400pt}}
\multiput(1007.00,406.93)(0.950,-0.485){11}{\rule{0.843pt}{0.117pt}}
\multiput(1007.00,407.17)(11.251,-7.000){2}{\rule{0.421pt}{0.400pt}}
\multiput(1020.00,399.93)(0.874,-0.485){11}{\rule{0.786pt}{0.117pt}}
\multiput(1020.00,400.17)(10.369,-7.000){2}{\rule{0.393pt}{0.400pt}}
\multiput(1032.00,392.93)(0.950,-0.485){11}{\rule{0.843pt}{0.117pt}}
\multiput(1032.00,393.17)(11.251,-7.000){2}{\rule{0.421pt}{0.400pt}}
\multiput(1045.00,385.93)(0.950,-0.485){11}{\rule{0.843pt}{0.117pt}}
\multiput(1045.00,386.17)(11.251,-7.000){2}{\rule{0.421pt}{0.400pt}}
\multiput(1058.00,378.93)(0.874,-0.485){11}{\rule{0.786pt}{0.117pt}}
\multiput(1058.00,379.17)(10.369,-7.000){2}{\rule{0.393pt}{0.400pt}}
\multiput(1070.00,371.93)(0.950,-0.485){11}{\rule{0.843pt}{0.117pt}}
\multiput(1070.00,372.17)(11.251,-7.000){2}{\rule{0.421pt}{0.400pt}}
\multiput(1083.00,364.93)(1.123,-0.482){9}{\rule{0.967pt}{0.116pt}}
\multiput(1083.00,365.17)(10.994,-6.000){2}{\rule{0.483pt}{0.400pt}}
\multiput(1096.00,358.93)(0.950,-0.485){11}{\rule{0.843pt}{0.117pt}}
\multiput(1096.00,359.17)(11.251,-7.000){2}{\rule{0.421pt}{0.400pt}}
\multiput(1109.00,351.93)(0.874,-0.485){11}{\rule{0.786pt}{0.117pt}}
\multiput(1109.00,352.17)(10.369,-7.000){2}{\rule{0.393pt}{0.400pt}}
\multiput(1121.00,344.93)(0.950,-0.485){11}{\rule{0.843pt}{0.117pt}}
\multiput(1121.00,345.17)(11.251,-7.000){2}{\rule{0.421pt}{0.400pt}}
\multiput(1134.00,337.93)(0.950,-0.485){11}{\rule{0.843pt}{0.117pt}}
\multiput(1134.00,338.17)(11.251,-7.000){2}{\rule{0.421pt}{0.400pt}}
\multiput(1147.00,330.93)(0.874,-0.485){11}{\rule{0.786pt}{0.117pt}}
\multiput(1147.00,331.17)(10.369,-7.000){2}{\rule{0.393pt}{0.400pt}}
\multiput(1159.00,323.93)(0.950,-0.485){11}{\rule{0.843pt}{0.117pt}}
\multiput(1159.00,324.17)(11.251,-7.000){2}{\rule{0.421pt}{0.400pt}}
\multiput(1172.00,316.93)(0.950,-0.485){11}{\rule{0.843pt}{0.117pt}}
\multiput(1172.00,317.17)(11.251,-7.000){2}{\rule{0.421pt}{0.400pt}}
\multiput(1185.00,309.93)(0.950,-0.485){11}{\rule{0.843pt}{0.117pt}}
\multiput(1185.00,310.17)(11.251,-7.000){2}{\rule{0.421pt}{0.400pt}}
\multiput(1198.00,302.93)(0.874,-0.485){11}{\rule{0.786pt}{0.117pt}}
\multiput(1198.00,303.17)(10.369,-7.000){2}{\rule{0.393pt}{0.400pt}}
\multiput(1210.00,295.93)(0.950,-0.485){11}{\rule{0.843pt}{0.117pt}}
\multiput(1210.00,296.17)(11.251,-7.000){2}{\rule{0.421pt}{0.400pt}}
\multiput(1223.00,288.93)(0.950,-0.485){11}{\rule{0.843pt}{0.117pt}}
\multiput(1223.00,289.17)(11.251,-7.000){2}{\rule{0.421pt}{0.400pt}}
\multiput(1236.00,281.93)(0.874,-0.485){11}{\rule{0.786pt}{0.117pt}}
\multiput(1236.00,282.17)(10.369,-7.000){2}{\rule{0.393pt}{0.400pt}}
\multiput(1248.00,274.93)(0.950,-0.485){11}{\rule{0.843pt}{0.117pt}}
\multiput(1248.00,275.17)(11.251,-7.000){2}{\rule{0.421pt}{0.400pt}}
\multiput(1261.00,267.93)(0.950,-0.485){11}{\rule{0.843pt}{0.117pt}}
\multiput(1261.00,268.17)(11.251,-7.000){2}{\rule{0.421pt}{0.400pt}}
\multiput(1274.00,260.93)(1.123,-0.482){9}{\rule{0.967pt}{0.116pt}}
\multiput(1274.00,261.17)(10.994,-6.000){2}{\rule{0.483pt}{0.400pt}}
\multiput(1287.00,254.93)(0.874,-0.485){11}{\rule{0.786pt}{0.117pt}}
\multiput(1287.00,255.17)(10.369,-7.000){2}{\rule{0.393pt}{0.400pt}}
\multiput(1299.00,247.93)(0.950,-0.485){11}{\rule{0.843pt}{0.117pt}}
\multiput(1299.00,248.17)(11.251,-7.000){2}{\rule{0.421pt}{0.400pt}}
\multiput(1312.00,240.93)(0.950,-0.485){11}{\rule{0.843pt}{0.117pt}}
\multiput(1312.00,241.17)(11.251,-7.000){2}{\rule{0.421pt}{0.400pt}}
\multiput(1325.00,233.93)(0.874,-0.485){11}{\rule{0.786pt}{0.117pt}}
\multiput(1325.00,234.17)(10.369,-7.000){2}{\rule{0.393pt}{0.400pt}}
\multiput(1337.00,226.93)(0.950,-0.485){11}{\rule{0.843pt}{0.117pt}}
\multiput(1337.00,227.17)(11.251,-7.000){2}{\rule{0.421pt}{0.400pt}}
\multiput(1350.00,219.93)(0.950,-0.485){11}{\rule{0.843pt}{0.117pt}}
\multiput(1350.00,220.17)(11.251,-7.000){2}{\rule{0.421pt}{0.400pt}}
\multiput(1363.00,212.93)(0.874,-0.485){11}{\rule{0.786pt}{0.117pt}}
\multiput(1363.00,213.17)(10.369,-7.000){2}{\rule{0.393pt}{0.400pt}}
\multiput(1375.00,205.93)(0.950,-0.485){11}{\rule{0.843pt}{0.117pt}}
\multiput(1375.00,206.17)(11.251,-7.000){2}{\rule{0.421pt}{0.400pt}}
\multiput(1388.00,198.93)(0.950,-0.485){11}{\rule{0.843pt}{0.117pt}}
\multiput(1388.00,199.17)(11.251,-7.000){2}{\rule{0.421pt}{0.400pt}}
\multiput(1401.00,191.93)(0.950,-0.485){11}{\rule{0.843pt}{0.117pt}}
\multiput(1401.00,192.17)(11.251,-7.000){2}{\rule{0.421pt}{0.400pt}}
\multiput(1414.00,184.93)(0.874,-0.485){11}{\rule{0.786pt}{0.117pt}}
\multiput(1414.00,185.17)(10.369,-7.000){2}{\rule{0.393pt}{0.400pt}}
\multiput(1426.00,177.93)(0.950,-0.485){11}{\rule{0.843pt}{0.117pt}}
\multiput(1426.00,178.17)(11.251,-7.000){2}{\rule{0.421pt}{0.400pt}}
\end{picture}